\documentclass[10pt]{article}

\usepackage{amsmath} 
\usepackage{graphicx}
\usepackage{amssymb}
\usepackage{epstopdf}
\usepackage[square,authoryear]{natbib}
\newcommand{\tb}{\mbox{\boldmath$\hat \varphi$}}
\newcommand{\pb}{\mbox{\boldmath$\hat \theta$}}

\title{Thin elastic shells with variable thickness\\
for lithospheric flexure of one-plate planets}
\author{Mikael Beuthe}
\date{\it Royal Observatory of Belgium, Brussels, Belgium. E-mail: mbeuthe@oma.be}

\begin{document}

\maketitle

\begin{abstract}
Planetary topography can either be modeled as a load supported by the lithosphere, or as a dynamical effect due to lithospheric flexure caused by mantle convection.
In both cases the response of the lithosphere to external forces can be calculated with the theory of thin elastic plates or shells.
On one-plate planets the spherical geometry of the lithospheric shell plays an important role in the flexure mechanism.
So far the equations governing the deformations and stresses of a spherical shell have only been derived under the assumption of a shell of constant thickness.
However local studies of gravity and topography data suggest large variations in the thickness of the lithosphere.
In this article we obtain the scalar flexure equations governing the deformations of a thin spherical shell with variable thickness or variable Young's modulus.
The resulting equations can be solved in succession, except for a system of two simultaneous equations, the solutions of which are the transverse deflection and an associated stress function.
In order to include bottom loading generated by mantle convection, we extend the method of stress functions to include loads with a toroidal tangential component.
We further show that toroidal tangential displacement always occurs if the shell thickness varies, even in the absence of toroidal loads.
We finally prove that the degree-one harmonic components of the transverse deflection and of the toroidal tangential displacement are independent of the elastic properties of the shell and are associated with translational and rotational freedom.
While being constrained by the static assumption, degree-one loads can deform the shell and generate stresses.
The flexure equations for a shell of variable thickness are useful not only for the prediction of the gravity signal in local admittance studies, but also for the construction of stress maps in tectonic analysis. 
\end{abstract}


\section{Introduction}

Terrestrial planets are flattened spheres only at first sight: close-up views reveal a rich topography with unique characteristics for each planet.
Unity in diversity is found by studying the support mechanism for topographic deviations from the hydrostatic planetary shape.
A simple mechanism, called isostasy, postulates mountains floating with iceberg-like roots in the higher density mantle.
Another simple mechanism assumes that mountains stand on a rigid membrane called the {\it mechanical lithosphere}.
The two simple models predict very different gravity signals, from very weak in the first to very strong in the second.
The truth lies in-between: an encompassing model views topographic structures as loads on an elastic shell with finite rigidity, called the {\it elastic lithosphere} (the elastic lithosphere is a subset of the mechanical lithosphere).
The rigidity depends on the elastic properties of the rocks and on the {\it apparent elastic thickness} of the lithosphere.
The latter parameter is the main objective of many studies, since its value can be related to the lithospheric composition and temperature.
Another important application of the model of lithospheric flexure is the determination of stress maps, which can then be compared with the observed distribution of tectonic features.

Besides the important assumption of elasticity, the model of lithospheric flexure is often simplified by two approximations.
The first one states that the area to be analyzed is sufficiently small so that the curved lithospheric shell can be modeled as a flat plate.
The second approximation states that the shell (or plate) is thin, which means that deformations are small and that the elastic thickness is small with respect to the wavelength of the load. 

On Earth, the model of lithospheric flexure has been very successful for the understanding of the topography of the oceanic plates, whereas the analysis of continental plates is fraught with difficulties due to the very old and complex structure of the continents.
As far as we know, plate tectonics do not occur at the present time on other terrestrial planets, which are deemed {\it one-plate} planets \citep{solomon78}, although the term {\it single-shell} would fit better because of the curvature.
A single shell can support loads of much larger extent and greater weight, the best example of which is the huge Tharsis volcanic formation covering a large portion of Mars.
Indeed such a load cannot be supported by the bending moments present in a thin flat plate, whereas it can be supported by stresses tangent to the shell: the shell acts as a membrane.

The first application of thin shell theory to a planet was done by \citet{brotchie69} for the lithosphere of the Earth (see also \citet{brotchie71}).
However the approximation of flat plate theory was seen to be sufficient when it was understood that the lithosphere of the Earth is broken into several plates \citep{tanimoto98}.
\citet{brotchie69} do not consider tangential loads and their flexure equation only includes dominant terms in derivatives; their equation is thus a special case of the equations of \citet{kraus} and \citet{vlasov} discussed below.
The articles reviewed hereafter use the thin shell theory of Kraus or Vlasov unless mentioned otherwise.

On the Moon, \citet{solomon79} used Brotchie's equation to study displacement and stress in mare basins.
\citet{turcotte81} estimated gravity-topography ratios for the mascons and discussed the type of stress supporting topographic loads.
\citet{arkanihamed98} modeled the support of mascons with Brotchie's equation.
\citet{sugano04} and \citet{crosby05} estimated the elastic thickness of the lithosphere from Lunar Prospector data.

On Mars, the dominance of the Tharsis rise in the topography led to numerous applications of the theory of thin elastic shells to lithospheric flexure.
\citet{thurber78}, \citet{comer85}, \citet{hall86} and \citet{janle86} used Brotchie's equation to estimate the lithospheric thickness under Martian volcanoes.
\citet{turcotte81} studied the transition between bending and membrane regimes.
\citet{willemann82} analyzed the lithospheric support of Tharsis.
\citet{sleep85} analyzed the membrane stress distribution on the whole surface.
\citet{banerdt92}, \citet{banerdt00} and \citet{phillips01} used a model of lithospheric flexure including membrane support, bending stresses and tangent loads \citep{banerdt86} in order to study the global stress distribution.
\citet{arkanihamed00} determined the elastic thickness beneath large volcanoes with Brotchie's equation whereas
\citet{johnson00} estimated the elastic thickness beneath the North Polar Cap.
Using local admittance analysis with spatiospectral methods, \citet{mcgovern02} determined the elastic thickness at various locations \citep[see also][]{mcgovern04}.
\citet{mckenzie02} made local admittance analyses of line-of-sight gravity data both with flat plates and with spherical shell models.
\citet{turcotte02} used the spherical shell formula for a one-dimensional wavelet analysis of the admittance in order to determine the average elastic thickness of the lithosphere.
\citet{zhong03} and \citet{lowry03} studied the support of the Tharsis rise with an hybrid model including the flexure of a thin elastic shell as well as the internal loading of a thermal plume in the mantle.
\citet{belleguic05} determined the elastic thickness and the density beneath large volcanoes.
\citet{searls06} investigated the elastic thickness and the density beneath the Utopia and Hellas basins.

Venus is considered as a one-plate planet but does not have giant volcanic or tectonic structures comparable to Tharsis.
The spherical shell model has thus not been used as often for Venus as for Mars.
\citet{banerdt86} studied the global stress distribution.
\citet{janle88} and \citet{johnson94} used Brotchie's equation to estimate the lithospheric thickness in various locations.
\citet{sandwell97} computed global strain trajectories for comparison with observed tectonics.
\citet{lawrence03} inverted the admittance in order to estimate the elastic thickness and mantle density anomalies over two lowland regions and one volcanic rise.
\citet{anderson06} established a global map of the elastic thickness based on local admittance analysis.
Mercury's topography and gravity fields are not yet known well enough to warrant the application of a thin elastic shell model.
We refer to \citet{wieczorek06} for a review of recent results regarding the lithosphere of terrestrial planets.

The elastic thickness of the lithosphere is not at all homogeneous over the surface of a planet.
For example, \citet{mcgovern04} (for Mars) and \citet{anderson06} (for Venus) find lithospheric thickness variations of more than 100~km.
The former study explains the variation in lithospheric thickness in terms of different epochs of loading, as the lithosphere is thickening with time. Other studies however inferred that spatial variations in lithospheric thickness on Mars are as important as temporal variations \citep{solomon82,solomon90,comer85}.
Thin spherical shell models have always been applied with a constant elastic thickness for the whole lithosphere.
Local studies are done by windowing both the data (gravity and topography) and the model predictions for gravity \citep{simons97,wieczorek05}.
The assumptions behind these methods are that the elastic thickness is constant within the window and that the area outside the window can be neglected.
The first assumption is of course true for a small enough window, but the size of the window is limited from below by the resolution of the data \citep{wieczorek05}.
Even if the first assumption were true, the second assumption is violated in two ways (unless the elastic thickness is spatially constant).
First, the deformation of the shell within the window as well as the associated stress field are both modified if the elastic thickness is changed in the area outside the window.
Second, the value of the predicted gravity field within the window depends on the shell deflection outside the window.

These reasons make it interesting to develop a model of the lithospheric flexure for a spherical shell of variable thickness.
Although a full inversion of the gravity and topography data is impractical with such a model because of the huge size of the parameter space, other applications are of high interest. 
For example a two-stage inversion can be considered: in the first stage a constant elastic thickness is assumed, and the resulting values are used in the second stage as a starting point for an inversion with variable elastic thickness (the parameter space can also be constrained with an a priori).
Moreover this model can be used to produce synthetic data and thus allows us to check the validity of inversions assuming a constant elastic thickness.
Finally, stress and strain fields can be computed for given variations of the elastic thickness, with the aim of comparing stress and strain maps with tectonic features.

General equations governing the deformations of a thin elastic shell have been given by various authors \citep[e.g.][]{love,vlasov,kraus}.
However the possibility of variable shell thickness is only considered at the early stage where the strain-displacement relationships, Hooke's law and the equilibrium equations are separately derived for the thin shell.
The combination of these three sets of equations into a unique equation for the transverse deflection is made under the restriction of constant elastic thickness, {\it `owing to the analytical complications which would otherwise arise'} \citep[see][p.~199]{kraus}.

This article is dedicated to the derivation of the minimum set of equations governing the deformations of a thin elastic spherical shell with variable thickness.
Using Kraus' method of stress functions, we find that the transverse deflection is the solution of a simultaneous system of two differential equations of the fourth order.
Contrary to the case of constant thickness, these equations cannot be combined due to the presence of products of derivatives of the thickness and derivatives of the deflection or of the stress function.
In order to include bottom loading generated by mantle convection, we extend the method of stress functions to include toroidal tangential loads, which were not considered by \citet{kraus}.
Non-toroidal loading is for example generated by tangential lithostatic forces whereas toroidal loading could be due to mantle flow generating drag at the base of the lithosphere (mantle flow also produces non-toroidal loading).
With applications to tectonics in mind, we derive the equations relating the tangential displacements and the stresses to the transverse deflection and the stress functions.
We further show that toroidal tangential displacement occurs even if there is no toroidal loading (unless the shell thickness is constant).
Finally we prove three properties specific to the degree-one harmonic components:
(1) the degree-one transverse deflection and the degree-one toroidal tangential displacement drop from the elasticity equations because they represent rigid displacements of the whole shell, (2) the transverse and tangential components of degree-one loads are related so that the shell does not accelerate, (3) degree-one loads can deform the shell and generate stresses.
Though our aim is to introduce a variable shell thickness, the final equations are also valid for a variable Young's modulus (Poisson's ratio must be kept constant).

Another way to take into account variations of the lithospheric thickness consists in treating the lithosphere as a three-dimensional spherical solid that is elastic \citep{metivier06} or viscoelastic \citep[e.g.][]{zhongpau03,latychev05,wang06}.
The resulting equations are exact (no thin shell approximation) and can be solved with finite element methods.
The thin shell assumption is probably satisfied for known planetary lithospheres; in case of doubt, it is advisable to compare the results of thin shell theory with three-dimensional models assuming constant elastic thickness: static thick shell models \citep[e.g.][]{banerdt82,janes90,arkanihamed02} or time-dependent viscoelastic models \citep[e.g.][]{zhong00,zhong02}.
The advantage of thin shell equations is their two-dimensional character, making them much easier to program and quicker to solve on a computer.
Solving faster either gives access to finer two-dimensional grids or allows to examine a larger parameter space.

We choose to work within the formalism of differential calculus on curved surfaces, without which the final equations would be cumbersome.
Actually the only tool used in this formalism is the covariant derivative, which can be seen by geophysicists as just a way of combining several terms into one `derivative'.
All necessary formulas are given in the Appendix.
The possibility of writing a differential equation in terms of covariant derivatives (in a tensorial form) also provides a consistency check.
The presence of derivatives that cannot be included into covariant derivatives is simply forbidden.
This is not without meaning for differential equations of the fourth order including products of derivatives.

In section \ref{fundamental}, we show how to obtain the strain-displacement relationships, Hooke's law and the equilibrium equations for a thin spherical shell.
These equations are available in the literature for the general case of a thin shell \citep[e.g.][]{kraus}, but we derive them for the spherical case in a simpler way, starting directly with the metric for the spherical shell.
We examine in detail the various approximations made to obtain the thin shell theory of flexure, refraining until the end from taking the `thin shell' quantitative limit in order to ascertain its influence on the final equations.
In section \ref{resolution}, we use the method of stress functions to obtain the flexure equations governing the displacements and the stresses.
In section \ref{properties}, we give the final form of the flexure equations in the thin shell approximation.
We also study the covariance and the degree-one projection of the flexure equations.
In section \ref{limitcases}, we examine various limit cases in which the flexure equations take a simpler form: the membrane limit, the Euclidean limit and the limit of constant thickness.


\section{Fundamental equations of elasticity}
\label{fundamental}

\subsection{Three-dimensional elasticity theory}
\label{3Dtheory}

Linear elasticity theory is based on three sets of equations. We directly state them in tensorial form for an isotropic material, since they are derived in Cartesian coordinates in many books \citep[e.g.][]{ranalli,synge}.
Recall that, in tensorial notation, there is an implicit summation on indices that are repeated on the same side of an equation.
The first set of equations includes strain-displacement relationships:
\begin{equation}
\epsilon_{ij} = \frac{1}{2} \left( u_{i,j} + u_{j,i} \right) \, ,
\label{strain}
\end{equation}
where $\epsilon_{ij}$ is the infinitesimal {\it strain tensor} and $u_i$ are the finite {\it displacements}.
The `comma' notation denotes the spatial derivative (see Appendix \ref{Aderivatives}).

The second set includes the constitutive equations of elasticity, or Hooke's law, relating the strain tensor and the {\it stress tensor} $\sigma_{ij}$:
\begin{equation}
\sigma_{ij} = \lambda \, \epsilon \, \delta_{ij} + 2 \, G \, \epsilon_{ij} \, ,
\label{hooke}
\end{equation}
where $\epsilon=\epsilon_{11}+\epsilon_{22}+\epsilon_{33}$ and $\delta_{ij}=1$ if $i\!=\!j$, otherwise it equals zero.
The parameter $\lambda$ is known as the {\it first Lam\' e constant}.
The parameter $G$ is known as the {\it second Lam\' e constant}, or the {\it shear modulus}, or the {\it modulus of rigidity}.
Boundary conditions are given by
\begin{equation}
\sigma_{ij} \, n_j=T_i \, ,
\end{equation}
with $n_j$ being the normal unit vector of the surface element and $T_i$ being the surface force per unit area. 

The third set includes equations of motion which reduce to equilibrium equations for stresses if the problem is static:
\begin{equation}
\sigma_{ij,j} = 0 \, .
\label{stress}
\end{equation}
Body forces, such as gravity, are assumed to be absent.
Both strain and stress tensors are symmetric: $\epsilon_{ij}=\epsilon_{ji}$ and $\sigma_{ij}=\sigma_{ji}$.

These three-dimensional equations do not yet have the right form for the description of the deformations of a two-dimensional spherical shell.
Various methods have been used to generate appropriate equations.
\citet{love} and \citet{timoshenkoW} derive strain-displacements and equilibrium equations directly on the surface of the sphere (the latter only for the special cases of no bending or axisymmetrical loading).
\citet{sokolnikoff} derives strain-displacement equations in three-dimensional curvilinear coordinates using an arbitrary diagonal metric but states without proof the equilibrium equations in curvilinear coordinates.
\citet{kraus} uses Sokolnikoff's form of strain-displacement equations and derives equilibrium equations for an arbitrary two-dimensional surface using Hamilton's principle (i.e. virtual displacements).

Instead of directly deriving equations on the two-dimensional surface of the sphere, we will first obtain their form in three-dimensional curvilinear coordinates and then restrict them to the surface of the sphere.
The first step can elegantly be done through the use of tensors \citep{synge}.
Equations (\ref{strain})-(\ref{stress}) are tensorial with respect to orthogonal transformations, but not with respect to other coordinate transforms (one reason being the presence of usual derivatives).
In other words, they are only valid in Cartesian coordinates.
In a three-dimensional Euclidean space, tensorial equations have a simplified form in Cartesian coordinates because supplementary terms that make them tensorial with respect to arbitrary coordinate transformations are zero.
The missing terms can be reconstructed by using a set of rules, such as the replacement of usual derivatives by covariant derivatives and the substitution of tensorial contraction to sum on components.
Correspondence rules lead to the following three sets of equations:
\begin{eqnarray}
\epsilon_{ij} &=& \frac{1}{2} \left( u_{i|j} + u_{j|i} \right) \, ,
\label{straincurvi} \\
\sigma_{ij} &=& \lambda \, \epsilon \, g_{ij} + 2 \, G \, \epsilon_{ij} \, ,
\label{hookecurvi}
\\
g^{jk} \, \sigma_{ij|k} &=& 0 \, ,
\label{stresscurvi}
\end{eqnarray}
where $\epsilon=g^{kl} \, \epsilon_{kl}$. The notation $u_{i|j}$ denotes the covariant derivative of $u_i$ (see Appendix \ref{Aderivatives}).
The metric and its inverse are noted $g_{ij}$ and $g^{ij}$, respectively.
Tensorial components cannot be expressed in a normalized basis (except for Cartesian coordinates) which is more common for physical interpretation (see Appendix \ref{Acomponents}). Covariant components in equations (\ref{straincurvi})-(\ref{stresscurvi}) are related to components defined in a normalized basis (written with a hat) by:
\begin{eqnarray}
u_i &=& \sqrt{g_{ii}} \; \hat u_i \, ,
\nonumber \\
\epsilon_{ij} &=& \sqrt{g_{ii}g_{jj}} \; \hat \epsilon_{ij} \, ,
\nonumber \\
\sigma_{ij} &=& \sqrt{g_{ii}g_{jj}} \; \hat \sigma_{ij} \, ,
\nonumber
\end{eqnarray}
where there is no implicit summation on repeated indices.

In the next section we will introduce additional assumptions in order to restrict the equations to the two-dimensional surface of a spherical shell.

\subsection{Spherical shell}
\label{sphericalshell}

\subsubsection{Assumptions of the thin shell theory}

Suppose that the two first coordinates are the colatitude $\theta$ and longitude $\varphi$ on the surface of the sphere, whereas the third coordinate $\zeta$ is radial. $R$ is the shell radius.
 
Assumptions of the thin shell theory are \citep[see][chap.~2.2]{kraus}:
\begin{enumerate}
\item The shell is thin (say less than one tenth of the radius of the sphere).
\item The deflections of the shell are small.
\item The transverse normal stress is negligible: $\sigma_{\zeta\zeta}=0$.
\item Normals to the reference surface of the shell remain normal to it and undergo no change of length during deformation:
$\epsilon_{\theta\zeta}=\epsilon_{\varphi\zeta}=\epsilon_{\zeta\zeta}=0$.
\end{enumerate}
The second assumption allows us to use linear equations to describe the deflections.
The third and fourth assumptions are not fully consistent: we refer to \citet{kraus} for more details.
We will relax them in the derivation of the equations for the deflection of a spherical shell.
The crucial assumption is $\sigma_{\zeta\zeta}=0$ which is essential for the restriction of Hooke's law to the two-dimensional shell.
As we will see later, $\sigma_{\zeta\zeta}$ cannot be zero since it is related to the non-zero transverse load (besides the fact that it is incompatible with a vanishing transverse strain).
What is absolutely necessary is that $\sigma_{\zeta\zeta}\ll\sigma_{ii}$ for $i=(\theta,\varphi)$.
In section \ref{breakdown}, we will show that this condition is satisfied if the wavelength of the load is much larger than the thickness of the shell.

The reference surface is the middle surface of the shell. With the aim of integrating out the third coordinate, a coordinate system is chosen so that the radial coordinate $\zeta$ is zero on the reference surface. The metric is given by
\begin{equation}
ds^2 = \left( R + \zeta \right)^2 \left( d \theta^2 + \sin^2 \theta \, d \varphi^2 \right) + d \zeta^2 \, .
\label{shellmetric}
\end{equation}
Christoffel symbols necessary for the computation of the covariant derivatives are given in Appendix \ref{A3Dgeom}.

\subsubsection{Strain-displacement equations}

With the metric (\ref{shellmetric}), the strain-displacement equations (\ref{straincurvi}) become
\begin{eqnarray}
\hat \epsilon_{\theta\theta} &=& \frac{1}{R+\zeta} \, ( \hat u_{\theta , \theta} + \hat u_\zeta ) \, ,
\nonumber \\
\hat \epsilon_{\varphi\varphi} &=& \frac{1}{R+\zeta} \, ( \csc \theta \, \hat u_{\varphi , \varphi} + \cot \theta \, \hat u_\theta + \hat u_\zeta ) \, ,
\nonumber \\
\hat \epsilon_{\theta\varphi} &=& \frac{1}{2} \, \frac{1}{R+\zeta} \, ( \csc \theta \, \hat u_{\theta , \varphi} - \cot \theta \, \hat u_\varphi + \hat u_{\varphi , \theta} ) \, ,
\label{strainsphere1} \\
\hat \epsilon_{\zeta\zeta} &=& \hat u_{\zeta , \zeta} \, ,
\nonumber \\
\hat \epsilon_{\theta\zeta} &=& \frac{1}{2} \, \frac{1}{R+\zeta} \, \left( (R+\zeta) \, \hat u_{\theta , \zeta} - \hat u_\theta + \hat u_{\zeta , \theta} \right) \, ,
\nonumber \\
\hat \epsilon_{\varphi\zeta} &=& \frac{1}{2} \, \frac{1}{R+\zeta} \, \left( (R+\zeta) \, \hat u_{\varphi , \zeta} - \hat u_\varphi + \csc \theta \, \hat u_{\zeta , \varphi} \right) \, ,
\nonumber
\end{eqnarray}
where all quantities are given in a normalized basis.
The fourth assumption of the thin shell theory implies that the displacements are linearly distributed across the thickness of the shell, with the transverse displacement being constant. Displacements can thus be expanded to first order in $\zeta$:
\begin{equation}
(\hat u_\theta,\hat u_\varphi,\hat u_\zeta) = (v_\theta+\zeta \beta_\theta,v_\varphi+\zeta \beta_\varphi,w) \, .
\label{linearexp1}
\end{equation}
The coefficients $(v_\theta,v_\varphi,w,\beta_\theta,\beta_\varphi)$ are independent of $\zeta$: $(v_\theta,v_\varphi,w)$ represent the components of the displacement vector of a point on the reference surface, whereas $(\beta_\theta,\beta_\varphi)$ represent the rotations of tangents to the reference surface oriented along the tangent axes.
We determine $\beta_\theta$ and $\beta_\varphi$ by applying the fourth assumption of the thin shell theory and the expansion (\ref{linearexp1}) to the last two strain-displacement equations (\ref{strainsphere1}):
\begin{eqnarray}
\beta_\theta &=& \frac{1}{R} \, ( v_\theta - w_{, \theta} ) \, ,
\nonumber \\
\beta_\varphi &=& \frac{1}{R} \, ( v_\varphi - \csc \theta \, w_{, \varphi} ) \, .
\label{linearexp2}
\end{eqnarray}
The substitution of the expansion (\ref{linearexp1}) into the fourth strain-displacement equation ($\hat\epsilon_{\zeta\zeta}=\hat u_{\zeta,\zeta}$) leads to the condition $\hat \epsilon_{\zeta\zeta} =0$, as postulated by the thin shell theory.

After the insertion of the expansion formulas (\ref{linearexp1}) and (\ref{linearexp2}), the first three strain-displacement equations (\ref{strainsphere1}) become
\begin{eqnarray}
\hat \epsilon_{\theta\theta} &=& \epsilon_\theta^0  + \frac{\zeta}{1+\zeta/R} \, \kappa_\theta^0 \, ,
\nonumber \\
\hat \epsilon_{\varphi\varphi} &=& \epsilon_\varphi^0 + \frac{\zeta}{1+\zeta/R} \, \kappa_\varphi^0 \, ,
\label{linearexp3} \\
2 \, \hat \epsilon_{\theta\varphi} &=& \gamma_{\theta\varphi}^0 +  \frac{\zeta}{1+\zeta/R} \, \tau^0 \, .
\nonumber
\end{eqnarray}
The extensional strains $\epsilon_\theta^0$, $\epsilon_\varphi^0$ and $\gamma_{\theta\varphi}^0$ are defined by
\begin{eqnarray}
\epsilon_\theta^0 &=& \frac{1}{R} \, ( v_{\theta , \theta} + w ) \, ,
\nonumber \\
\epsilon_\varphi^0 &=& \frac{1}{R} \, ( \csc \theta \, v_{\varphi , \varphi} + \cot \theta \, v_\theta + w ) \, ,
\label{strainsphere2}\\
\gamma_{\theta\varphi}^0 &=& \frac{1}{R} \, ( v_{\varphi ,\theta} - \cot \theta \, v_\varphi + \csc \theta \, v_{\theta , \varphi} ) \, .
\nonumber
\end{eqnarray}
They represent the normal and shearing strains of the reference surface.
The flexural strains $\kappa_\theta^0$, $\kappa_\varphi^0$ and $\tau^0$ are given by
\begin{eqnarray}
\kappa_\theta^0 &=& - \frac{1}{R^2} \, {\cal O}_1 \, w \, ,
\nonumber \\
\kappa_\varphi^0 &=& - \frac{1}{R^2} \, {\cal O}_2 \, w \, ,
\label{curvetwist} \\
\tau^0 &=&  - \frac{2}{R^2} \, {\cal O}_3 \, w \, .
\nonumber
\end{eqnarray}
They represent the changes in curvature and the torsion of the reference surface during deformation \citep{kraus}.
The differential operators ${\cal O}_{1,2,3}$ are defined by
\begin{eqnarray}
{\cal O}_1 &=& \frac{\partial^2}{\partial \theta^2} + 1 \, ,
\nonumber \\
{\cal O}_2 &=&  \csc^2 \theta \, \frac{\partial^2}{\partial \varphi^2} + \cot \theta \, \frac{\partial}{\partial \theta} + 1 \, ,
\label{opO} \\
{\cal O}_3 &=& \csc \theta  \left( \frac{\partial^2}{\partial \theta \partial \varphi} - \cot \theta \, \frac{\partial}{\partial \varphi} \right) \, .
\nonumber
\end{eqnarray}
The zero upper index in $(\epsilon_\theta^0,\epsilon_\varphi^0,\gamma_{\theta\varphi}^0,\kappa_\theta^0,\kappa_\varphi^0,\tau^0)$ refers to the reference surface and appears in order to follow Kraus' notation.

In Love's theory, one makes the approximation $\frac{1}{R+\zeta}\cong\frac{1}{R}$ in equations (\ref{linearexp3}). However this does not simplify the calculations when the shell is a sphere because it has only one radius of curvature (for a shell with two radii of curvature, this approximation is a great simplification).
Our choice to keep the factor $\frac{1}{R+\zeta}$ leads to the same results as the theory of Fl\"ugge-Lur'e-Byrne, explained in \citet[][chap.~3.3a]{kraus} or \citet[][p.~53]{novozhilov}, in which this factor is expanded up to second order.
In any case the choice between the approximation or the expansion of this factor does not affect the equations (derived in the next section) relating the stress and moment resultants to the strains: they are the same if $\frac{1}{R+\zeta}$ is approximated to zeroth order, expanded to second order or fully kept.

\subsubsection{Hooke's law}

When the metric is diagonal and the basis is normalized, Hooke's law (\ref{hookecurvi}) becomes
\begin{eqnarray}
\hat \sigma_{ii} &=& \lambda \sum_{k=1}^3 \hat  \epsilon_{kk} + 2 \, G \,  \hat  \epsilon_{ii}
\hspace{1cm} (i=j) \, ,
\nonumber \\
\hat \sigma_{ij} &=& 2 \, G \, \hat  \epsilon_{ij}
\hspace{1cm} (i \neq j) \, .
\nonumber
\end{eqnarray}
There is no implicit summation on repeated indices.

The third assumption of the thin shell theory, $\hat\sigma_{\zeta\zeta}=0$, can be used to eliminate $\hat\epsilon_{\zeta\zeta}$ from Hooke's law:
\begin{eqnarray}
\hat \sigma_{\theta\theta} &=& \frac{E}{1-\nu^2} \, ( \hat \epsilon_{\theta\theta} + \nu \, \hat \epsilon_{\varphi\varphi} ) \, ,
\nonumber \\
\hat \sigma_{\varphi\varphi} &=& \frac{E}{1-\nu^2} \, ( \hat \epsilon_{\varphi\varphi} + \nu \, \hat \epsilon_{\theta\theta} ) \, .
\nonumber
\end{eqnarray}
Young's modulus $E$ and Poisson's ratio $\nu$ are related to Lam\'e parameters by
\begin{eqnarray}
\lambda &=& \frac{E\nu}{(1+\nu)(1-2\nu)} \, ,
\nonumber \\
G&=& \frac{E}{2(1+\nu)} \, .
\nonumber
\end{eqnarray}
In principle, the fourth assumption of the thin shell theory, $\hat\epsilon_{\theta\zeta}=\hat\epsilon_{\varphi\zeta}=0$, leads to $\hat\sigma_{\theta\zeta}=\hat\sigma_{\varphi\zeta}=0$ but non-vanishing values must be retained for purposes of equilibrium.

The substitution of the expansion (\ref{linearexp3}) into the thin shell approximation of Hooke's law gives
\begin{eqnarray}
\hat \sigma_{\theta\theta} &=& \frac{E}{1-\nu^2} \left( \epsilon_\theta^0 + \nu \epsilon_\varphi^0
+ \frac{\zeta}{1+\zeta/R} \left( \kappa_\theta^0 + \nu \, \kappa_\varphi^0 \right) \right) \, ,
\nonumber \\
\hat \sigma_{\varphi\varphi} &=& \frac{E}{1-\nu^2} \left( \epsilon_\varphi^0 + \nu \epsilon_\theta^0
+ \frac{\zeta}{1+\zeta/R} \left( \kappa_\varphi^0 + \nu \, \kappa_\theta^0 \right) \right) \, ,
\label{linearexp4} \\
\hat \sigma_{\theta\varphi} &=& \frac{E}{2(1+\nu)} \left( \gamma_{\theta\varphi}^0
+ \frac{\zeta}{1+\zeta/R} \, \tau^0 \right) \, ,
\nonumber
\end{eqnarray}
with $\epsilon_\theta^0$, $\epsilon_\varphi^0$, $\gamma_{\theta\varphi}^0$, $\kappa_\theta^0$, $\kappa_\varphi^0$ and $\tau^0$ defined by equations (\ref{strainsphere2}) and (\ref{curvetwist}).
The expressions for $\hat\sigma_{\theta\zeta}$ and $\hat\sigma_{\varphi\zeta}$ will not be needed.

\begin{figure}
   \centering
   \includegraphics[width=8cm]{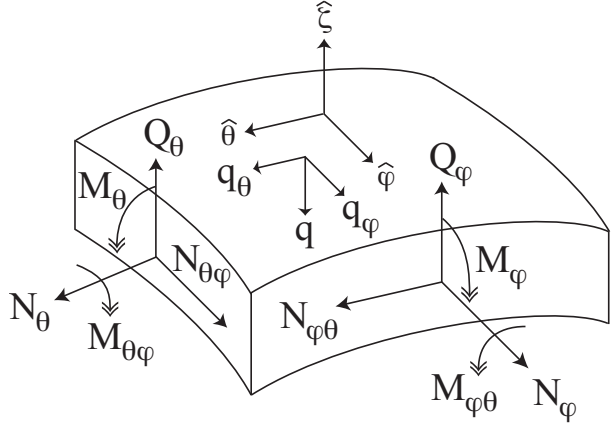} 
   \caption{Stress resultants and stress couples acting on a small element of the shell. The directions of the stress resultants (simple arrows) and the rotation sense of the stress couples (double arrows) correspond to positive components (tensile stress is positive). Loads $(q_\theta,q_\varphi,q)$ act on the reference surface.}
   \label{fig1}
\end{figure}

We now integrate the stress distributions across the thickness $h$ of the shell (see Figure \ref{fig1}).
The stress resultants and couples obtained in this way are defined per unit of arc length on the reference surface:
\begin{eqnarray}
N_i &=& \int_{-h/2}^{h/2} \hat \sigma_{ii} \, (1+\zeta/R) \, d\zeta \hspace{1cm} (i=\theta,\varphi) \, , 
\nonumber \\
N_{\theta\varphi} &=& N_{\varphi\theta} \;=\; \int_{-h/2}^{h/2} \hat \sigma_{\theta\varphi} \, (1+\zeta/R) \, d\zeta \, , 
\nonumber \\
Q_i &=& \int_{-h/2}^{h/2} \hat \sigma_{i\zeta} \, (1+\zeta/R) \, d\zeta \hspace{1cm} (i=\theta,\varphi) \, , 
\label{resultants1} \\
M_i &=& \int_{-h/2}^{h/2} \hat \sigma_{ii} \, (1+\zeta/R) \, \zeta \, d\zeta \hspace{1cm} (i=\theta,\varphi) \, , 
\nonumber \\
M_{\theta\varphi} &=& M_{\varphi\theta} \;=\; \int_{-h/2}^{h/2} \hat \sigma_{\theta\varphi} \, (1+\zeta/R) \, \zeta \, d\zeta \, .
\nonumber
\end{eqnarray}
We evaluate these integrals with the expansion (\ref{linearexp4}).
The tangential stress resultants are
\begin{eqnarray}
N_\theta &=& K \left(\epsilon_\theta^0 + \nu \, \epsilon_\varphi^0 \right) \, ,
\nonumber \\
N_\varphi &=& K \left(\epsilon_\varphi^0 + \nu \, \epsilon_\theta^0 \right) \, ,
\label{stressres} \\
N_{\theta\varphi} &=& K \, \frac{1-\nu}{2} \, \gamma_{\theta\varphi}^0 \, .
\nonumber
\end{eqnarray}
Explicit expressions for the transverse shearing stress resultants $Q_i$ are not needed since these quantities will be determined from the equilibrium equations.
The moment resultants are
\begin{eqnarray}
M_\theta &=& D \left( \kappa_\theta^0 + \nu \, \kappa_\varphi^0  + \frac{1}{R} \, \left(\epsilon_\theta^0 + \nu \, \epsilon_\varphi^0 \right) \right)\, ,
\nonumber \\
M_\varphi &=& D \left( \kappa_\varphi^0 + \nu \, \kappa_\theta^0 + \frac{1}{R} \, \left(\epsilon_\varphi^0 + \nu \, \epsilon_\theta^0 \right) \right)\, ,
\label{momentres} \\
M_{\theta\varphi} &=& D \, \frac{1-\nu}{2} \left( \tau^0 + \frac{1}{R} \, \gamma_{\theta\varphi}^0 \right) \, .
\nonumber
\end{eqnarray}
The {\it extensional rigidity} $K$ and the {\it bending rigidity} $D$ are defined by
\begin{eqnarray}
K &=& \frac{E h}{1-\nu^2} \, ,
\label{defK} \\
D &=& \frac{E h^3}{12 (1-\nu^2)} \, .
\label{defD}
\end{eqnarray}
Their dimensionless ratio $\xi$ is a large number,
\begin{equation}
\xi = R^2 \, \frac{K}{D} = \frac{12 R^2}{h^2} \, ,
\label{defxi}
\end{equation}
the inverse of which will serve as an expansion parameter for thin shell theory.

\subsubsection{Equilibrium equations}

With the metric (\ref{shellmetric}), the components $\theta$, $\varphi$ and $\zeta$ of the equilibrium equations (\ref{stresscurvi}) respectively become
\begin{eqnarray}
(R+\zeta) \left( \left( \sin \theta \, \hat \sigma_{\theta\theta} \right)_{, \theta} + \hat \sigma_{\theta\varphi , \varphi} - \cos \theta \, \hat \sigma_{\varphi\varphi}
+ \sin \theta \, \hat \sigma_{\zeta\theta} \right) + \sin \theta \left( \left( R+\zeta \right)^2 \hat \sigma_{\zeta\theta} \right)_{, \zeta} &=& 0 \, ,
\nonumber \\
(R+\zeta) \left( \left( \sin \theta \, \hat \sigma_{\theta\varphi} \right)_{, \theta} + \hat \sigma_{\varphi\varphi , \varphi} + \cos \theta \, \hat \sigma_{\theta\varphi}
+ \sin \theta \, \hat \sigma_{\zeta\varphi} \right) + \sin \theta \left( \left( R + \zeta \right)^2 \hat \sigma_{\zeta\varphi} \right)_{, \zeta} &=& 0 \, ,
\nonumber \\
(R+\zeta) \left( \left( \sin \theta \, \hat \sigma_{\theta\zeta} \right)_{, \theta} + \hat \sigma_{\varphi\zeta , \varphi} - \sin \theta \, \left( \hat \sigma_{\theta\theta} + \hat \sigma_{\varphi\varphi} \right) \right)
+ \sin \theta \left( \left( R + \zeta \right)^2 \hat \sigma_{\zeta\zeta} \right)_{, \zeta} &=& 0 \, ,
\nonumber
\end{eqnarray}
where the equations have been multiplied by $\sin \theta \, (R+\zeta)$, $(R+\zeta)$ and $\sin \theta \, (R+\zeta)^2$, respectively.
The stress components are given in a normalized basis.

The integration on $\zeta$ of these three equations in the range $[-h/2,h/2]$ yields the equilibrium equations for the forces:
\begin{eqnarray}
\left( \sin \theta \, N_\theta \right)_{,\theta} + N_{\theta\varphi ,\varphi} - \cos \theta \, N_\varphi + \sin \theta \, Q_\theta + R \, \sin \theta \, q_\theta &=&  0 \, ,
\label{equil1} \\
\left( \sin \theta N_{\theta\varphi} \right)_{, \theta} + N_{\varphi , \varphi} + \cos \theta \, N_{\theta\varphi} + \sin \theta \, Q_\varphi + R \, \sin \theta \, q_\varphi &=& 0 \, ,
\label{equil2} \\
\left( \sin \theta \, Q_\theta \right)_{, \theta} + Q_{\varphi , \varphi} - \sin \theta \, \left( N_\theta + N_\varphi \right) - R \, \sin \theta \, q &=& 0 \, ,
\label{equil3}
\end{eqnarray}
where $q_\theta$ and $q_\varphi$ are the components of the tangential load vector per unit area of the reference surface:
\[
\left[ \left( R + \zeta \right)^2 \hat \sigma_{\zeta i} \right]_{-h/2}^{h/2} = R^2 \, q_i \hspace{1cm} (i=\theta,\varphi) \, .
\]
We choose the convention that tensile stresses are positive (see Figure \ref{fig1}).
The transverse load per unit area of the reference surface is noted $q$ and is taken to be positive toward the center of the sphere:
\begin{equation}
\left[ \left( R + \zeta \right)^2 \hat \sigma_{\zeta\zeta} \right]_{-h/2}^{h/2} = - R^2 \, q \, .
\label{defq}
\end{equation}

The first two equilibrium equations for the stresses can also be multiplied by $\zeta$ before the integration to yield the equilibrium equations for the moments:
\begin{eqnarray}
\left( \sin \theta \, M_\theta \right)_{, \theta} + M_{\theta\varphi , \varphi} - \cos \theta \, M_\varphi - R \, \sin \theta \, Q_\theta &=& 0 \, ,
\label{moment1}
\\
\left( \sin \theta \, M_{\theta\varphi} \right)_{, \theta} + M_{\varphi , \varphi} + \cos \theta \, M_{\theta\varphi} - R \, \sin \theta \, Q_\varphi &=& 0 \, .
\label{moment2}
\end{eqnarray}
We have neglected small terms in $\left[ \zeta \left( R + \zeta \right)^2 \hat \sigma_{\zeta i} \right]_{-h/2}^{h/2}$ where $i=(\theta,\varphi)$.
A third equilibrium equation for the moments exists but has the form of an identity: $M_{\theta\varphi}=M_{\varphi\theta}$.


\section{Resolution}
\label{resolution}

\subsection{Available methods}

At this stage the elastic theory for a thin spherical shell involves 17 equations: six strain-displacement relationships (\ref{strainsphere2})-{\ref{curvetwist}), six stress-strain relations (\ref{stressres})-(\ref{momentres}) making Hooke's law, and five equilibrium equations (\ref{equil1})-(\ref{equil1}) and (\ref{moment1})-(\ref{moment2}).
There are 17 unknowns: six strain components $(\epsilon_\theta^0,\epsilon_\varphi^0,\gamma_{\theta\varphi}^0,\kappa^0_\theta,\kappa^0_\varphi,\tau^0)$, three displacements $(w,v_\theta,v_\varphi)$, three tangential stress resultants $(N_\theta,N_\varphi,N_{\theta\varphi})$, two transverse shearing stress resultants $(Q_\theta,Q_\varphi)$, and three moment resultants $(M_\theta,M_\varphi,M_{\theta\varphi})$.
The three equations (\ref{linearexp4}) are also needed if the tangential stresses $(\hat\sigma_{\theta\theta},\hat\sigma_{\theta\varphi},\hat\sigma_{\varphi\varphi})$ are required.
The quantities of primary interest to us are the transverse deflection and the tangential stresses (sometimes tangential strain is preferred, as in \citet{sandwell97} or \citet{banerdt00}).
We thus want to find the minimum set of equations that must be solved to determine these quantities.

We are aware of two methods of resolution \citep[p.~66]{novozhilov}. In the first one, we insert the strain-displacement relationships into Hooke's law, and substitute in turn Hooke's law into the equilibrium equations. This method yields three simultaneous differential equations for the displacements.
Once the displacements are known, it is possible to compute the strains and the stresses.
The second method supplements the equilibrium equations with the {\it equations of compatibility} \citep[p.~27]{novozhilov} that relate the partial derivatives of the strain components.
It is then convenient to introduce the so-called {\it stress functions} \citep[p.~243]{kraus}, without direct physical interpretation, which serve to define the stress resultants without introducing the tangential displacements.
Equations relating the transverse displacement and the stress functions are then found by applying the third equation of equilibrium and the third equation of compatibility (it is also possible to use all three equations of compatibility in order to directly solve for the stress and moment resultants).
Once the transverse displacement and the stress functions are known, stresses can be computed.

If the shell thickness is constant, the deformations of a thin spherical shell can be completely calculated with both methods. If the shell thickness is variable, the three equations governing displacements, obtained with the first method, cannot be decoupled and are not easy to solve. Kraus' method with stress functions leads to a system of three equations (relating the transverse displacement and the two stress functions), in which the first equation is decoupled and solved before the other two. This method thus provides a system of equations much easier to solve and will be chosen in this article.

When solving the equations, one usually assumes from the beginning the large $\xi$ limit, i.e. $1+\xi\cong\xi$ where $\xi$ is defined by equation (\ref{defxi}).
We will only take this limit at the end of the resolution.
This procedure will not complicate the computations, since we have to compute anyway many new terms because of the variable shell thickness.

\subsection{Differential operators}
\label{diffop}

We will repeatedly encounter the operators ${\cal O}_i$ which intervene in the expressions (\ref{curvetwist}) for the flexural strains $(\kappa^0_\theta,\kappa^0_\varphi,\tau^0)$.
Since we are looking for scalar equations, we need to find out how the operators ${\cal O}_i$ can be combined in order to yield scalar expressions, i.e. expressions that are invariant with respect to changes of coordinates on the sphere.

The first thing is to relate the ${\cal O}_i$ to tensorial operators.
Starting from the covariant derivatives on the sphere $\nabla_i$, we construct the following tensorial differential operators of the second degree in derivatives:
\begin{equation}
{\cal D}_{ij} = \nabla_i \nabla_j + g_{ij} \, ,
\label{tensorop}
\end{equation}
where $\nabla_i$ denotes the covariant derivative (see Appendix \ref{Aderivatives}).
These operators give zero when applied on spherical harmonics of degree one (considered as scalars):
\begin{equation}
{\cal D}_{ij} \, Y_{1m} = 0 \hspace{1cm} (m=-1,0,1) \, .
\label{propertydeg1}
\end{equation}
This property can be explicitly checked on the spherical harmonics (\ref{SH1}) with the metric and the formulas for the double covariant derivatives given in Appendix \ref{A2Dgeom}.

In two-dimensional spherical coordinates ($\theta,\varphi$), the three operators ${\cal O}_i$ defined by equations (\ref{opO}) are related to the operators ${\cal D}_{ij}$ acting on a scalar function $f$ through
\begin{eqnarray}
{\cal O}_1 \, f &=& {\cal D}_{\theta\theta} \, f \, ,
\nonumber \\
{\cal O}_2 \, f &=& \csc^2 \theta \, {\cal D}_{\varphi\varphi} \, f \, ,
\label{opObis} \\
{\cal O}_3 \, f &=& \csc \theta \, {\cal D}_{\theta\varphi} \, f \, .
\nonumber
\end{eqnarray}
The operators ${\cal O}_{1,2,3}f$ actually correspond to normalized ${\cal D}_{ij}f$, i.e. ${\cal D}_{ij}f/( \sqrt{ g_{ii} g_{jj} } )$.

The usual derivatives of the operators ${\cal O}_i$ satisfy the useful identities (\ref{id1})-(\ref{id2}) which are proven in Appendix \ref{Aidentities}.
These identities are the consequence of the path dependence of the parallel transport of vectors on the curved surface of the sphere.

Invariant expressions are built by contracting all indices of the differential operators in their tensorial form.
The indices can be contracted with the inverse metric $g^{ij}$ or with the antisymmetric tensor $\varepsilon^{ij}$ (see Appendix \ref{A2Dgeom}), which should not be confused with the strain tensor $\epsilon_{ij}$.
In the following, $a$ and $b$ are scalar functions on the sphere.

At degree 2, the only non-zero contraction of the ${\cal D}_{ij}$ is related to the Laplacian (\ref{laplacian}):
\begin{eqnarray}
\Delta' a
&\equiv& g^{ij} \, {\cal D}_{ij} \, a
\nonumber \\
&=& \left( \Delta + 2 \right) a
\label{deltaprime} \\
&=&   \left( {\cal O}_1 + {\cal O}_2 \right) a
\nonumber \\
&=&
a_{,\theta , \theta} +  \cot \theta \, a_{, \theta} + \csc^2 \theta \, a_{, \varphi , \varphi} + 2 \, a \, .
\nonumber
\end{eqnarray}

At degree 4, a scalar expression symmetric in $(a,b)$ is given by
\begin{eqnarray}
{\cal A}(a\,;b)
&\equiv& [\Delta' a] [\Delta' b] - [{\cal D}_{ij} \, a] [{\cal D}^{ij} \, b] \, ,
\nonumber \\
&=& [\Delta \, a][\Delta \, b] - [\nabla_i \nabla_j \, a][\nabla^i \nabla^j \, b] + [\Delta \, a] \, b + a \, [\Delta \, b] + 2 \, a \, b
\label{defA} \\
&=& [{\cal O}_1 \, a][{\cal O}_2 \, b] +  [{\cal O}_2 \, a][{\cal O}_1 \, b] - 2 \, [{\cal O}_3 \, a][{\cal O}_3 \, b]
\nonumber \\
&=&
   \left( a_{,\theta , \theta} + a \right) \left( \csc^2 \theta \, b_{, \varphi , \varphi} + \cot \theta \, b_{, \theta} + b \right)
+ \left( \csc^2 \theta \, a_{, \varphi , \varphi} + \cot \theta \, a_{, \theta} + a \right)  \left( b_{,\theta , \theta} + b \right)
\nonumber \\ &&
- 2 \csc^2 \theta \left( a_{, \theta , \varphi} - \cot \theta \, a_{, \varphi} \right) \left( b_{, \theta , \varphi} - \cot \theta \, b_{, \varphi} \right) \, .
\nonumber
\end{eqnarray}
where upper indices are raised with the inverse metric: ${\cal D}^{ij}=g^{ik}g^{jl}{\cal D}_{kl}$.
The action of an operator does not extend beyond the brackets enclosing it.

If $a$ is constant, ${\cal A}(a\,;b)=a\,\Delta'b$.
It is useful to define an associated operator ${\cal A}_0$ that gives zero if its first argument is constant:
\begin{equation}
{\cal A}_0(a\,;b) = {\cal A}(a\,;b) - a \left[ \Delta' b \right] \, .
\label{defA0}
\end{equation}

A scalar expression of degree 4 antisymmetric in $(a,b)$ is given by
\begin{eqnarray}
{\cal B}_1(a\,;b)
&\equiv& g^{ij} \, \varepsilon^{kl} \left[ {\cal D}_{ik} \, a \right] \left[ {\cal D}_{jl} \, b \right]
\nonumber \\
&=& g^{ij} \, \varepsilon^{kl} \left[ \nabla_i \nabla_k \, a \right] \left[ \nabla_j \nabla_l \, b \right]
\label{defB1} \\
&=& \left[ \left( {\cal O}_1 - {\cal O}_2 \right) a \right] \left[ {\cal O}_3 b \right]
                       - \left[ {\cal O}_3 a \right]  \left[ \left( {\cal O}_1 - {\cal O}_2 \right) b \right]
\nonumber \\
&=& \csc \theta \left( a_{,\theta , \theta} - \csc^2 \theta \, a_{, \varphi , \varphi} - \cot \theta \, a_{, \theta} \right) \left( b_{, \theta , \varphi} - \cot \theta \, b_{, \varphi} \right)
\nonumber \\ &&
- \csc \theta \left( a_{, \theta , \varphi} - \cot \theta \, a_{, \varphi} \right) \left( b_{,\theta , \theta} - \csc^2 \theta \, b_{, \varphi , \varphi} - \cot \theta \, b_{, \theta} \right)  .
\nonumber
\end{eqnarray}

We will also need another operator of degree 4:
\begin{eqnarray}
{\cal B}_2(a\,;b)
&\equiv& \varepsilon^{ij} \left[ \nabla_i a \, \right] \left[\nabla_j \, \Delta' b \right]
\nonumber \\
&=&  \csc \theta \left( a_{, \theta} \left[ \Delta' b \right]_{, \varphi} - a_{, \varphi} \left[ \Delta' b \right]_{, \theta} \right) \, .
\label{defB2}
\end{eqnarray}
The sum of the operators ${\cal B}_1$ and ${\cal B}_2$ is noted ${\cal B}$:
\begin{equation}
{\cal B}(a\,;b) = {\cal B}_1(a\,;b) + {\cal B}_2(a\,;b) \, .
\label{defB}
\end{equation}
If either $a$ or $b$ is constant, ${\cal B}(a\,;b)=0$.

The operators ${\cal A}$ and ${\cal B}$ have an interesting property proven in Appendix \ref{Adegree1}: for arbitrary scalar functions $a$ and $b$, ${\cal A}(a\,;b)$ and ${\cal B}(a\,;b)$ do not have a degree-one term in their spherical harmonic expansion.
This is not true of ${\cal A}_0$, ${\cal B}_1$ and ${\cal B}_2$.

\subsection{Transverse displacement}
\label{transverse}

\subsubsection{Resolution of the equations of equilibrium}
\label{transverse1}

The first step consists in finding expressions for the moment resultants $(M_\theta,M_\varphi,M_{\theta\varphi})$ in terms of the transverse displacement and the stress resultants.
The extensional strains $(\epsilon_\theta^0,\epsilon_\varphi^0,\gamma_{\theta\varphi}^0)$ can be eliminated from the equations for stress and moment resultants (\ref{stressres})-(\ref{momentres}).
The flexural strains $(\kappa_\theta^0,\kappa_\varphi^0,\tau^0)$ depend on the transverse displacement $w$ through equations (\ref{curvetwist}).
We thus obtain
\begin{eqnarray}
M_\theta &=& - \frac{D}{R^2} \, ( {\cal O}_1 + \nu \, {\cal O}_2 ) \, w + \frac{R}{\xi} \, N_\theta \, ,
\nonumber \\
M_\varphi &=& - \frac{D}{R^2} \, ( {\cal O}_2 + \nu \, {\cal O}_1 ) \, w + \frac{R}{\xi} \, N_\varphi \, ,
\label{moment3} \\
M_{\theta\varphi} &=& - \frac{D}{R^2} \, (1-\nu) \, {\cal O}_3 \, w + \frac{R}{\xi} \, N_{\theta\varphi} \, ,
\nonumber
\end{eqnarray}
where the parameter $\xi$ is defined by equation (\ref{defxi}).

The second step consists in solving the equilibrium equations for moments in order to find the transverse shearing stress resultants $(Q_\theta,Q_\varphi)$.
We substitute expressions (\ref{moment3}) into equations (\ref{moment1})-(\ref{moment2}).
Knowing that the stress resultants satisfy the equilibrium equations (\ref{equil1})-(\ref{equil2}), we obtain new expressions for $Q_\theta$ and $Q_\varphi$ (identities (\ref{id1})-(\ref{id2}) are helpful):
\begin{eqnarray}
Q_\theta &=& - \, \frac{\eta}{R^3} \, \left( D \, \Delta' w \right)_{, \theta} - (1-\eta) \, R \, q_\theta
                   \nonumber \\ &  & 
          + \,  \frac{\eta}{R^3} \, (1-\nu) \, \left( D_{, \theta} \, {\cal O}_2 - \csc \theta \, D_{, \varphi} \, {\cal O}_3 \right) \, w
                   - \frac{1}{\eta} \left( \eta_{, \theta} \, N_\theta + \csc \theta \, \eta_{, \varphi} \, N_{\theta\varphi} \right) \, ,
\label{Qp} \\
Q_\varphi &=& - \, \frac{\eta}{R^3} \, \csc \theta \, \left( D \, \Delta' w \right)_{, \varphi} - (1-\eta) \, R \, q_\varphi
                   \nonumber \\ &  &
           +  \, \frac{\eta}{R^3} \, (1-\nu) \, \left( - D_{, \theta} \, {\cal O}_3 + \csc \theta \, D_{, \varphi} \, {\cal O}_1 \right) \, w
                   - \frac{1}{\eta} \left( \eta_{, \theta} \, N_{\theta\varphi} + \csc \theta \, \eta_{, \varphi} \, N_\varphi \right) \, .
\label{Qt}
\end{eqnarray}
The operator $\Delta'$ is defined by equation (\ref{deltaprime}) and $\eta$ is a parameter close to 1:
\begin{equation}
\eta \equiv \frac{\xi}{1+\xi} = \left( 1+\frac{h^2}{12R^2} \right)^{-1} \, .
\end{equation}

The third step consists in finding expressions for the stress resultants $(N_\theta,N_\varphi,N_{\theta\varphi})$ in terms of stress functions by solving the first two equilibrium equations (\ref{equil1})-(\ref{equil2}).
Let us define the following linear combinations of the stress and moment resultants:
\begin{equation}
\left( P_\theta,P_\varphi,P_{\theta\varphi} \right) = \left( N_\theta + \frac{1}{R}M_\theta,N_\varphi + \frac{1}{R}M_\varphi,N_{\theta\varphi} + \frac{1}{R}M_{\theta\varphi} \right) \, .
\label{linearcombi}
\end{equation}
We observe that these linear combinations satisfy simplified equations of equilibrium:
\begin{eqnarray}
\left( \sin \theta \, P_\theta \right)_{,\theta} + P_{\theta\varphi ,\varphi} - \cos \theta \, P_\varphi + R \, \sin \theta \, q_\theta &=&  0 \, ,
\nonumber \\
\left( \sin \theta \, P_{\theta\varphi} \right)_{, \theta} + P_{\varphi , \varphi} + \cos \theta \, P_{\theta\varphi} + R \, \sin \theta \, q_\varphi &=& 0 \, .
\label{equilP}
\end{eqnarray}

Comparing these equilibrium equations with the identities (\ref{id1})-(\ref{id2}), we see that the homogeneous equations (i.e. equations (\ref{equilP}) with a zero tangential load $q_\theta=q_\varphi=0$) are always satisfied if
\begin{equation}
\left( P_\theta,P_\varphi,P_{\theta\varphi} \right) = \left( {\cal O}_2,{\cal O}_1,-{\cal O}_3 \right) F \, ,
\label{homog}
\end{equation}
where $F$ is an auxiliary function called {\it stress function}.
For the moment, this function is completely arbitrary apart from being scalar and differentiable.

Particular solutions of the full equations (\ref{equilP}) can be found if we express the tangential load ${\bf q}_T=q_\theta \pb+q_\varphi \tb$ in terms of the surface gradient of a scalar potential $\Omega$ ({\it consoidal} or {\it poloidal} component) and the surface curl of a vector potential $V{\bf\hat r}$ ({\it toroidal} component):
\begin{equation}
{\bf q}_T = - \frac{1}{R} \, {\bf \bar \nabla} \Omega + \frac{1}{R} \, {\bf \bar \nabla} \times (V{\bf\hat r}) \, .
\label{tangentload1}
\end{equation}
Surface operators are defined in Appendix \ref{Adiffop}, where the terms consoidal/poloidal are also discussed.
The covariant components of ${\bf q}_T$ are $(q_\theta,\sin\theta q_\varphi)$ and can be expressed as
$-\frac{1}{R}\Omega_{, i}+\frac{1}{R}g^{jk}\varepsilon_{ik}V_{, j}$, which gives
\begin{eqnarray}
q_\theta &=& - \frac{1}{R} \, \Omega_{, \theta} + \frac{1}{R \sin \theta} \, V_{, \varphi} \, ,
\nonumber \\
\sin \theta \, q_\varphi &=& - \frac{1}{R} \, \Omega_{, \varphi} - \frac{\sin \theta}{R} \, V_{, \theta} \, .
\label{tangentload2}
\end{eqnarray}
If the tangential load is consoidal ($V=0$), a particular solution of equations (\ref{equilP}) is given by 
\begin{equation}
\left( P_\theta,P_\varphi,P_{\theta\varphi} \right) = (1,1,0) \, \Omega \, .
\label{partic1}
\end{equation}
If the tangential load is toroidal ($\Omega=0$), a particular solution of equations (\ref{equilP}) is given by
\begin{equation}
\left( P_\theta,P_\varphi,P_{\theta\varphi} \right) = \left( 2 {\cal O}_3, - 2 {\cal O}_3, {\cal O}_2 - {\cal O}_1 \right) H \, ,
\label{partic2}
\end{equation}
where we have introduced a second stress function $H$ which satisfies the constraint
\begin{equation}
\Delta' H = - V + V_0  \, ,
\label{diffeq0}
\end{equation}
where $V_0$ is a constant (identities (\ref{id1})-(\ref{id2}) are useful).
This equation allows us to determine the stress function $H$ if the toroidal source $V$ is known.

The general solution of the equations (\ref{equilP}) is given by the sum of the general solution (\ref{homog}) of the homogeneous equations and the two particular solutions (\ref{partic1})-(\ref{partic2}) of the full equations:
\begin{equation}
\left( P_\theta,P_\varphi,P_{\theta\varphi} \right) =  \left( {\cal O}_2 F + \Omega + 2 {\cal O}_3 H, {\cal O}_1 F + \Omega - 2 {\cal O}_3 H,-{\cal O}_3 F + \left( {\cal O}_2 - {\cal O}_1 \right) H \right) \, .
\label{eqPplus}
\end{equation}
The stress resultants $( N_\theta,N_\varphi,N_{\theta\varphi} )$ can now be obtained from $(P_\theta,P_\varphi,P_{\theta\varphi})$ by using equations (\ref{linearcombi}) and (\ref{moment3}):
\begin{equation}
\left( N_\theta,N_\varphi,N_{\theta\varphi} \right) = \eta \left( P_\theta,P_\varphi,P_{\theta\varphi} \right) + \eta \frac{D}{R^3} \left( {\cal O}_1+\nu{\cal O}_2,{\cal O}_2+\nu{\cal O}_1,\left( 1-\nu \right) {\cal O}_3 \right) w \, ,
\end{equation}
which finally give
\begin{eqnarray}
N_\theta &=& \eta \left( {\cal O}_2 \, F + \frac{D}{R^3} \left( \Delta' - \left( 1-\nu \right) {\cal O}_2 \right) w + \Omega  + 2 \, {\cal O}_3 H \right) \, ,
\nonumber \\
N_\varphi &=&  \eta \left( {\cal O}_1 \, F + \frac{D}{R^3} \left( \Delta' - \left( 1-\nu \right) {\cal O}_1 \right) w + \Omega  - 2 \, {\cal O}_3 H \right) \, ,
\label{eqN}
\\
N_{\theta\varphi} &=& \eta \left( - {\cal O}_3 \, F  + (1-\nu) \, \frac{D}{R^3} \, {\cal O}_3 \, w +  \left( {\cal O}_2 - {\cal O}_1 \right) H \right) \, .
\nonumber 
\end{eqnarray}

The fourth step consists in expressing the third equation of equilibrium (\ref{equil3}) in terms of the transverse displacement $w$ and the stress functions $F$ and $H$.
For this purpose, it is handy to express the transverse shearing stress resultants $(Q_\theta,Q_\varphi)$ in terms of $(w,F,H)$.
We thus substitute $N_\varphi$, $N_\theta$, $N_{\theta\varphi}$, given by equations (\ref{eqN}), into the expressions for $Q_\theta$ and $Q_\varphi$, given by equations (\ref{Qp})-(\ref{Qt}):
\begin{eqnarray}
Q_\theta &=& - \, \frac{1}{R^3} \, \left( \eta D \, \Delta' w \right)_{, \theta}
               +  \frac{1-\nu}{R^3} \, \left( \left(\eta D \right)_{, \theta} {\cal O}_2 - \csc \theta \left(\eta D \right)_{, \varphi} {\cal O}_3 \right) w
                   \nonumber \\ &  &
                  - \left( \eta_{, \theta} \, {\cal O}_2 - \csc \theta \, \eta_{, \varphi} \, {\cal O}_3 \right) F
                  + \, \Omega_{, \theta} - (\eta \Omega)_{, \theta} - \csc \theta \, V_{, \varphi}
                  \nonumber \\ &  &
                - \, 2 \left( \eta_{, \theta} \, {\cal O}_3 + \csc \theta \, \eta_{, \varphi} \, {\cal O}_2  \right) H
               + \csc \theta \left( \eta_{, \varphi} \, \Delta' H - \eta (\Delta' H)_{, \varphi} \right)
                  \, ,
\label{Qpbis} \\
Q_\varphi &=& - \, \frac{1}{R^3} \, \csc \theta \, \left( \eta D \, \Delta' w \right)_{, \varphi}
               + \frac{1-\nu}{R^3} \, \left( - \left(\eta D \right)_{, \theta} {\cal O}_3 + \csc \theta \left(\eta D \right)_{, \varphi} {\cal O}_1 \right) w
                   \nonumber \\ &  &
               - \left( - \eta_{, \theta} \, {\cal O}_3 + \csc \theta \, \eta_{, \varphi} \, {\cal O}_1 \right) F
              + \csc \theta \, ( \Omega_{, \varphi} - (\eta \Omega)_{, \varphi} ) + V_{, \theta}
               \nonumber \\ &  &
               + \, 2 \left( \eta_{, \theta} \, {\cal O}_1 + \csc \theta \, \eta_{, \varphi} \, {\cal O}_3 \right) H
                - \left( \eta_{, \theta} \, \Delta' H - \eta (\Delta' H)_{, \theta} \right)
                  \, .
\label{Qtbis}
\end{eqnarray}

The various terms present in $Q_\theta$ and $Q_\varphi$ can be classified into generic types according to their differential structure.
Each generic type will contribute in a characteristic way to the third equation of equilibrium. It is worthwhile to compute the generic contribution of each type before using the full expressions of the stress resultants with their multiple terms. Identities (\ref{id1})-(\ref{id2}) are helpful in this calculation.
We thus write the third equation of equilibrium (\ref{equil3}) as follows:
\begin{equation}
{\cal I} \left( Q_\theta \,; Q_\varphi \,; 0 \right) - \left( N_\theta + N_\varphi \right)  - R \, q = 0 \, ,
\label{equil3bis}
\end{equation}
where the differential operator ${\cal I}$ is defined for arbitrary expressions $(X,Y,Z)$ by
\begin{equation}
{\cal I}(X\,;Y\,;Z) = \csc \theta \left( \sin \theta \, X \right)_{, \theta} + \csc \theta \, Y_{, \varphi} - \cot \theta \, Z \, .
\label{defI}
\end{equation}
This operator must be evaluated for the following generic types present in $(Q_\theta,Q_\varphi)$:
\begin{eqnarray}
{\cal I} \left( a_{, \theta} \,; \csc \theta \, a_{, \varphi} \,; 0 \right) &=& \Delta a \, ,
\nonumber \\
{\cal I} \left( -\csc \theta \, a_{, \varphi} \,; a_{, \theta} \,; 0 \right) &=& 0 \, ,
\nonumber \\
{\cal I} \left( a_{,\theta} \, {\cal O}_2 b - \csc\theta \, a_{,\varphi} \, {\cal O}_3 b \,; -a_{,\theta} \, {\cal O}_3 b + \csc\theta \, a_{,\varphi} \, {\cal O}_1 b \,; 0 \right) &=& {\cal A}_0(a\,;b) \, ,
\label{Xvalue}\\
{\cal I} \left( a_{, \theta} \, {\cal O}_3 b + \csc \theta \, a_{, \varphi} \, {\cal O}_2 b \,; - a_{, \theta} \, {\cal O}_1 b - \csc \theta \, a_{, \varphi} \, {\cal O}_3 b \,; 0 \right) &=& {\cal B}_1(a \,; b) \, ,
\nonumber \\
{\cal I} \left( - \csc \theta \left( a_{, \varphi} \, \Delta' b - a [\Delta' b]_{, \varphi} \right) ; a_{, \theta} \, \Delta' b - a [\Delta' b]_{, \theta} \,; 0 \right) &=& 2 \, {\cal B}_2(a \,; b) \, ,
\nonumber
\end{eqnarray}
where $a$ and $b$ are scalar functions.
The operators ${\cal A}_0$, ${\cal B}_1$ and ${\cal B}_2$ are defined in section \ref{diffop}.

We now substitute $(Q_\theta,Q_\varphi)$ and $(N_\varphi,N_\theta)$ into the third equation of equilibrium (\ref{equil3bis}) and use formulas (\ref{Xvalue}).
We thus obtain the first of the differential equations that relate $w$ and the stress functions $F$ and $H$:
\begin{equation}
\Delta' \left( \eta D \, \Delta' w \right) - (1-\nu) \, {\cal A}(\eta D \,; w) + R^3 \, {\cal A}(\eta \,; F) + 2 R^3 \, {\cal B}(\eta \,; H)
= -R^4 \, q + R^3 \left( \Delta \Omega - \Delta' (\eta \Omega) \right) \, .
\label{diffeq1}
\end{equation}
The operators $\Delta'$, ${\cal A}$ and ${\cal B}$ are defined in section \ref{diffop}.

\subsubsection{Compatibility relation}
\label{compatibility}

A second equation relating the transverse displacement and the stress functions comes from the compatibility relation
which is derived by eliminating $(v_\theta,v_\varphi)$ from the strain-displacement equations (\ref{strainsphere2}):
\begin{equation}
\left( \sin \theta \, \gamma_{\theta\varphi , \varphi}^0 \right)_{, \theta}
= \left( \sin^2 \theta \; \epsilon_{\varphi , \theta}^0 \right)_{, \theta} + \epsilon_{\theta , \varphi , \varphi}^0 - \sin \theta \, \cos \theta \; \epsilon_{\theta , \theta}^0
+ \, 2 \, \sin^2 \theta \; \epsilon_\theta^0  -  \frac{\sin^2 \theta}{R} \, \Delta' w \, .
\label{compatibility1}
\end{equation}
The strain components are related to the stress resultants through Hooke's law (\ref{stressres}).
The substitution of equations (\ref{stressres}) into the compatibility equation (\ref{compatibility1}) gives
\begin{equation}
\Delta' \left( \alpha \left( N_\theta + N_\varphi \right) \right)
- \frac{1}{R} \, \Delta' w
- (1+\nu) \, {\cal J} \left(  \alpha \, N_\theta \,;  \alpha \, N_\varphi \,;  \alpha \, N_{\theta\varphi} \right) = 0 \, .
\label{compatibility2}
\end{equation}
where $\alpha$ is the reciprocal of the extensional rigidity:
\begin{equation}
\alpha \equiv \frac{1}{K(1-\nu^2)} = \frac{1}{Eh} \, .
\label{defalpha}
\end{equation}
For arbitrary expressions $(X,Y,Z)$, ${\cal J}(X\,;Y\,;Z)$ is defined by
\begin{equation}
{\cal J} (X\,;Y\,;Z)  = 
\csc^2 \theta \left(
 \left( \sin^2 \theta \, X_{, \theta} \right)_{, \theta}
+ Y_{, \varphi , \varphi}
+ 2 \left( \sin \theta \, Z_{, \varphi} \right)_{, \theta}
\right)
- \cot \theta \, Y_{, \theta}
+ 2 \, Y
\, .
\end{equation}

As in the case of the third equation of equilibrium, it is practical to classify the terms present in $(N_\theta,N_\varphi,N_{\theta\varphi})$ into generic types and evaluate separately their contribution to the equation of compatibility.
There are three types of terms for which we must evaluate the operator ${\cal J}$ (identities (\ref{id1})-(\ref{id4}) are helpful in this calculation):
\begin{eqnarray}
{\cal J}(a\,;a\,;0) &=& \Delta' a \, ,
\nonumber \\
{\cal J}( a \, {\cal O}_2 b \,; a \, {\cal O}_1 b \,; - a \, {\cal O}_3 b ) &=& {\cal A}(a \,; b) \, ,
\label{Jeval1} \\
{\cal J} \left( 2 a \, {\cal O}_3 b \,; -2 a \, {\cal O}_3 b \,; a \, ( {\cal O}_2 - {\cal O}_1 ) b \right) &=& 2 \, {\cal B}(a \,; b) \, ,
\nonumber
\end{eqnarray}
where $a$ and $b$ are scalar functions.
The operators $\Delta'$, ${\cal A}$ and ${\cal B}$ are defined in section \ref{diffop}.

We now evaluate ${\cal J}$ in equation (\ref{compatibility2}) with expressions (\ref{eqN}) and formulas (\ref{Jeval1}):
\begin{eqnarray}
{\cal J} \left(  \alpha \, N_\theta \,;  \alpha \, N_\varphi \,;  \alpha \, N_{\theta\varphi} \right) &=&
{\cal A} (\eta\alpha \,; F) + \frac{1}{R^3} \, \Delta' \left( \eta\alpha D \, \Delta' w \right)
- \frac{1}{R^3} \, (1-\nu) \, {\cal A} (\eta\alpha D \,; w )
 \nonumber \\ &  &
+ \; \Delta' \left( \eta\alpha \, \Omega \right) + 2 \, {\cal B}(\eta\alpha \,; H) \, .
\label{Jeval2}
\end{eqnarray}

The term ${\cal A} (\eta\alpha D \,; w )$ can be rewritten with the help of the following equality:
\begin{equation}
\frac{1-\nu^2}{R^2} \, {\cal A} (\eta \alpha D \,; w ) = \Delta' w - {\cal A} (\eta \,; w ) \, ,
\label{relationA}
\end{equation}
since $(1-\nu^2)\eta\alpha D/R^2=\eta/\xi$ and $(\eta/\xi)_{, i}=-\eta_{, i}$.

We finally substitute $N_\theta+N_\varphi$, given by equations (\ref{eqN}), into the compatibility equation (\ref{compatibility2}) and use expressions (\ref{Jeval2})-(\ref{relationA}).
We thus obtain the second of the differential equations that relate $w$ and the stress functions $F$ and $H$:
\begin{equation}
\Delta' \left( \eta\alpha \, \Delta' F \right) - (1+\nu) \, {\cal A}(\eta\alpha \,; F) - \frac{1}{R} \, {\cal A}(\eta \,; w) - 2 (1+\nu) \, {\cal B}(\eta\alpha \,; H)
= - (1-\nu) \, \Delta' (\eta\alpha \, \Omega) \, .
\label{diffeq2}
\end{equation}

\subsection{Tangential displacements}
\label{tangential}

Assuming that the flexure equations (\ref{diffeq1}) and (\ref{diffeq2}) for the transverse displacement $w$ and the stress functions $(F,H)$ have been solved, we now show how to calculate the tangential displacements.

In analogy with the decomposition of the tangential load in equations (\ref{tangentload1})-(\ref{tangentload2}), the tangential displacement can be separated into consoidal and toroidal components:
\begin{equation}
{\bf v} = {\bf \bar \nabla} \, S + {\bf \bar \nabla} \times (T\,{\bf\hat r}) \, ,
\label{tangentdispl1}
\end{equation}
where $S$ and $T$ are the consoidal and toroidal scalars, respectively.
The covariant components of ${\bf v}$ are $(v_\theta,\sin \theta \, v_\varphi)$ and can be expressed as
$S_{, i}+g^{jk}\varepsilon_{ik}T_{, j}$ (see Appendix \ref{Adiffop}), which gives
\begin{eqnarray}
v_\theta &=& S_{, \theta} + \csc \theta \, \, T_{, \varphi} \, ,
\nonumber \\
\sin \theta \, v_\varphi &=& S_{, \varphi} - \sin \theta \, T_{, \theta} \, .
\label{tangentdispl2}
\end{eqnarray}

The strain-displacement equations (\ref{strainsphere2}) become
\begin{eqnarray}
\epsilon_\theta^0 &=& \frac{1}{R} \left( \left( {\cal O}_1 - 1 \right) S + {\cal O}_3 \, T + w \right) \, ,
\nonumber \\
\epsilon_\varphi^0 &=& \frac{1}{R} \left( \left( {\cal O}_2 - 1 \right) S - {\cal O}_3 \, T + w \right) \, ,
\label{strainsphere3}\\
\gamma_{\theta\varphi}^0 &=& \frac{1}{R} \left(  2 \, {\cal O}_3 S +  \left( {\cal O}_2 - {\cal O}_1 \right) T \right) \, .
\nonumber
\end{eqnarray}

The stress resultants $(N_\theta,N_\varphi,N_{\theta\varphi})$ given by equations (\ref{stressres}) become
\begin{eqnarray}
N_\theta &=& \frac{K}{R} \left( \Delta \, S + \left( 1 + \nu \right) w
              - (1-\nu) \left( \left( {\cal O}_2 -1 \right) S - {\cal O}_3 \, T \right) \right) \, ,
\nonumber \\
N_\varphi &=& \frac{K}{R} \left( \Delta \, S + \left( 1 + \nu \right) w
              - (1-\nu) \left( \left( {\cal O}_1 -1 \right) S + {\cal O}_3 \, T \right) \right) \, ,
\label{stressdispl} \\
N_{\theta\varphi} &=& \frac{K}{R} \, \frac{1-\nu}{2} \left( 2 \, {\cal O}_3 \, S + \left( {\cal O}_2 - {\cal O}_1 \right) T \right) \, .
\nonumber
\end{eqnarray}

The toroidal potential $T$ cancels in the sum $N_\theta+N_\varphi$:
\[
N_\theta+N_\varphi = \frac{K}{R} \, (1+\nu) \left( \Delta \, S + 2 w \right) \, .
\]
The consoidal displacement potential $S$ can thus be related to $(w,F,\Omega)$ by using expressions (\ref{eqN}) for the stress resultants:
\begin{equation}
\Delta \, S = R\eta\alpha \, (1-\nu) \left( \Delta' F + 2 \, \Omega \right) + \frac{\eta}{\xi} \, \Delta' w - 2 w \, ,
\label{diffeq3}
\end{equation}
where $\alpha$ is defined by equation (\ref{defalpha}).

It is more difficult to extract the toroidal displacement potential $T$.
When the shell thickness is constant, decoupled equations for the displacements can be found by suitable differentiation and combination of the three equilibrium equations (\ref{equil1})-(\ref{equil3}) for the stress resultants.
This method does not work if the shell thickness is variable because the resulting equations are coupled.
The trick consists in relating the tangential displacements to $(w,F,H)$ by the way of equations similar to the homogeneous part of the first two equilibrium equations, but with $(N_\theta,N_\varphi,N_{\theta\varphi})$ replaced by $\frac{R}{K}(N_\theta,N_\varphi,N_{\theta\varphi})$, so that derivatives of $\frac{1}{K}$ do not mix with derivatives of $T$.
We will thus calculate the following expression in two different ways (from equations (\ref{eqN}) and (\ref{stressdispl})):
\[
Z\left( \frac{R}{K} \, N_\theta \,; \frac{R}{K} \, N_\varphi \,; \frac{R}{K} \, N_{\theta\varphi} \right) = \csc^2 \theta \left( -X_{, \varphi}+\sin\theta\,Y_{,\theta} \right) \, ,
\]
where $X$ and $Y$ are defined by
\begin{eqnarray}
X &=& \sin \theta \; {\cal I}\left( \frac{R}{K} \, N_\theta \,; \frac{R}{K} \, N_{\theta\varphi} \,; \frac{R}{K} \, N_\varphi \right) \, ,
\nonumber \\
Y &=& \sin \theta \; {\cal I}\left( \frac{R}{K} \, N_{\theta\varphi} \,; \frac{R}{K} \, N_\varphi \,; -\frac{R}{K} \, N_{\theta\varphi} \right) \, .
\nonumber
\end{eqnarray}
The operator ${\cal I}$ is defined by equation (\ref{defI}).

As before, it is easier to begin with the evaluation of the operator $Z$ for generic contributions:
\begin{eqnarray}
Z ( a  \,; a \,; 0 ) &=& 0 \, ,
\nonumber \\
Z ( a {\cal O}_2 b \,; a {\cal O}_1 b \,; -a {\cal O}_3 b ) &=& - {\cal B}(a \,; b) \, ,
\nonumber \\
Z ( 2a {\cal O}_3 b \,; -2a {\cal O}_3 b \,; a ( {\cal O}_2 - {\cal O}_1 ) b ) &=& 2 \, {\cal A}(a \,; b) - \Delta' \left( a \Delta' b \right) \, ,
\nonumber
\end{eqnarray}
On the one hand, the evaluation $Z$ for $(N_\theta,N_\varphi,N_{\theta\varphi})$ given by equations (\ref{stressdispl}) gives
\[
- \frac{1-\nu}{2} \, \Delta \, \Delta' \, T \, .
\]
On the other hand, the evaluation of $Z$ for $(N_\theta,N_\varphi,N_{\theta\varphi})$ given by equations (\ref{eqN}) gives
\[
- R \, {\cal B} \left( \frac{\eta}{K} \,; F \right) + (1-\nu) \, {\cal B} \left( \frac{\eta}{\xi} \,; w \right)
+ 2 R \, {\cal A} \left( \frac{\eta}{K} \,; H \right) - R \, \Delta' \left( \frac{\eta}{K} \, \Delta' H \right) \, .
\]
The equality of the two previous formulas yields the sought equation for $T$:
\begin{equation}
\Delta \, \Delta' \, T =
2 R \, (1+\nu) \left( 
{\cal B} \left( \eta\alpha \,; F \right)  - 2 \, {\cal A} \left( \eta\alpha \,; H \right) + \Delta' \left( \eta\alpha \, \Delta' H \right)
\right)
+ 2 \, {\cal B} \left( \eta \,; w \right) \, .
\label{diffeq4}
\end{equation}
We have used the relation $(\eta/\xi)_{, i}=-\eta_{, i}$.

The equation for $T$ shows that toroidal displacement always occurs when the shell thickness varies.
The right-hand side of equation (\ref{diffeq4}) only vanishes when two conditions are met: (1) there is no toroidal source (so that the stress function $H$ vanishes) and (2) the shell thickness is constant (so that the terms in ${\cal B}$ vanish).

\subsection{Stresses}

Stresses can be computed from $(w,F,H)$ and $\Omega$ by substituting equations (\ref{curvetwist}), (\ref{stressres}) and (\ref{eqN}) into equations (\ref{linearexp4}):
\begin{eqnarray}
\hat \sigma_{\theta\theta} &=& \frac{\eta}{h} \left( {\cal O}_2 \, F + \Omega  + 2 \, {\cal O}_3 \, H \right)
+ \frac{E}{R(1-\nu^2)} \left( \frac{\eta}{\xi} - \frac{\zeta}{R+\zeta} \right) \left( \Delta' - \left( 1-\nu \right) {\cal O}_2 \right) w \, ,
\nonumber \\
\hat \sigma_{\varphi\varphi} &=& \frac{\eta}{h} \left( {\cal O}_1 \, F + \Omega  - 2 \, {\cal O}_3 \, H \right)
+ \frac{E}{R(1-\nu^2)} \left( \frac{\eta}{\xi} - \frac{\zeta}{R+\zeta} \right) \left( \Delta' - \left( 1-\nu \right) {\cal O}_1 \right) w \, ,
\label{stresses1}
\\
\hat \sigma_{\theta\varphi} &=& \frac{\eta}{h} \left( -{\cal O}_3 \, F + \left( {\cal O}_2 - {\cal O}_1 \right) H \right)
+ \frac{E}{R(1+\nu)} \left( \frac{\eta}{\xi} - \frac{\zeta}{R+\zeta} \right) {\cal O}_3 w \, .
\nonumber
\end{eqnarray}
Stresses at the surface are obtained by setting $\zeta=h/2$.


\section{Flexure equations and their properties}
\label{properties}

\subsection{Thin shell approximation}
\label{thinshelllimit}

The flexure equations derived in section \ref{resolution} already include several assumptions of the thin shell theory, but not yet the first one stating that the shell is thin.
Of course, the three other assumptions can be seen to be consequences of the first one \citep[see][chap.~2.2]{kraus}, but we have not imposed in a quantitative way the thinness condition on the equations.
We thus impose the limit of small $h/R$ or, equivalently, the limit of large $\xi$ (defined by equation (\ref{defxi})) on the flexure equations for $(H,F,w,S,T)$.
This procedure amounts to expand $\eta\approx1-1/\xi$ (neglecting terms in $1/\xi$ wherever appropriate) and to neglect the derivatives of $\eta$ in equations (\ref{diffeq0}), (\ref{diffeq1}), (\ref{diffeq2}), (\ref{diffeq3}) and (\ref{diffeq4}).
We thus obtain the final flexure equations for the displacements $(w,S,T)$ and for the stress functions $(F,H)$:
\begin{eqnarray}
\hspace{-1cm} &&
\Delta' H = - V + V_0  \, ,
\label{diffeqlim0} \\
\hspace{-1cm} &&
\Delta'  \left( D \, \Delta' w \right) - (1-\nu) \, {\cal A}(D \,; w) + R^3 \Delta' F
= -R^4 \, q - 2 \, R^3 \Omega + \frac{R^3}{\xi} \, \Delta \Omega \, ,
\label{diffeqlim1} \\
\hspace{-1cm} &&
\Delta' \left( \alpha \, \Delta' F \right) - (1+\nu) \, {\cal A}( \alpha \,; F) - \frac{1}{R} \; \Delta' w 
= - (1-\nu) \, \Delta' (\alpha \, \Omega) + 2 (1+\nu) \, {\cal B}(\alpha \,; H) \, ,
\label{diffeqlim2} \\
\hspace{-1cm} &&
\Delta \, S = R \alpha \, (1-\nu) \left( \Delta' F + 2 \, \Omega \right) + \frac{1}{\xi} \, \Delta' w - 2 w \, ,
\label{diffeqlim3} \\
\hspace{-1cm} &&
\Delta \, \Delta' \, T =
2 R \, (1+\nu) \left( {\cal B} \left( \alpha \,; F \right)  - 2 \, {\cal A} \left( \alpha \,; H \right) + \Delta' \left( \alpha \, \Delta' H \right) \right) \, .
\label{diffeqlim4}
\end{eqnarray}
Recall that the differential operators $\Delta'$, ${\cal A}$ and ${\cal B}$ are defined by equations (\ref{deltaprime}), (\ref{defA}) and (\ref{defB}).
The potential pairs $(\Omega,V)$ and $(S,T)$ are related to the tangential load and displacement by equations (\ref{tangentload1}) and (\ref{tangentdispl1}), respectively.
In the second equation, the term $R^3\Delta\Omega/\xi$ has been kept since it could be large if $\Omega$ has a short wavelength.
For the same reason, the term $\Delta'w/\xi$ has been kept in the fourth equation.

In the third and fifth equations, the terms depending on $H$ belong to the right-hand sides since they can be considered as a source once $H$ has been calculated from the first equation.
The same can be said of the terms depending on $w$ and $F$ in the fourth and fifth equations: they are supposed to be known from the simultaneous resolution of the second and third equations.
The difficulty in solving the equations thus lies with the two flexure equations for $w$ and $F$ which are linear with non-constant coefficients (all other equations are linear - in their unknowns - with constant coefficients). Once these two core equations have been solved, all other quantities are easily derived from them.

The bending rigidity $D$, defined by equation (\ref{defD}), characterizes the {\it bending regime}: the shell locally bends in a similar way as a flat plate undergoing small deflections with negligible stretching.
The parameter $\alpha$, defined by equation (\ref{defalpha}), is the reciprocal of the extensional rigidity $K$ and characterizes the {\it membrane regime} of the shell in which bending moments are negligible and the load is mainly supported by internal stresses tangent to the shell.
Since $D$ and $K$ are respectively proportional to the third and first power of the shell thickness, the membrane regime (in which the $D$-depending terms are neglected) is obtained in the limit of an extremely thin shell.
This observation and the fact that such a shell, lacking rigidity, cannot support bending moments justify the use of the term `membrane'.

The weight of the various terms in the equations depends on two competing factors: the magnitude of the coefficient multiplying the derivative and the number of derivatives. On the one hand, a coefficient containing $D$ will be smaller than a coefficient containing  $\alpha^{-1}$ since $\alpha\,D/R^2\sim\xi^{-1}$ is a small number (see equation (\ref{defxi})). On the other hand, a large number of derivatives will increase the weight of the term if the derived function has a small wavelength.
The transition between the membrane and the bending regimes thus depends on the wavelength of the load:
if the load has a large wavelength (with respect to the shell radius), the flexure of the shell will be well described by equations without the terms depending on $D$  (see section \ref{membranelimit}), whereas the flexure under loads of small wavelength is well described by equations keeping only the $D$-depending terms with the largest number of derivatives (see section \ref{euclideanlimit}).

Stress functions are associated with membrane stretching and give negligible contributions in the bending regime.
Formulas (\ref{eqN}) show that $F$ (respectively $H$) plays the role of potential for the stress resultants in the membrane regime when the load is transversal (respectively tangential toroidal). There is no stress function associated with the tangential consoidal component of the load because it is identical to $\Omega$, the consoidal potential of the load.

The variation of the shell thickness has two effects. First, it couples the spherical harmonic modes that are solutions to the equations for a shell of constant thickness. Second, the toroidal part remains intertwined with the transversal and consoidal parts, whereas it decouples if the thickness is constant. For example, the toroidal load is a source for the transverse deflection through the term ${\cal B}$ in the third flexure equation. Furthermore, the stress function $F$ is a source for the toroidal displacement in the fifth flexure equation.

For numerical computations, the flexure equations (\ref{diffeqlim0})-(\ref{diffeqlim4}) obtained in the thin shell approximation (to which we can add the equations (\ref{stresses1}) for the stresses) are adequate.
For this purpose, it is not useful to keep small terms in $1/\xi$ since the theory rests on assumptions only true for a thin shell.
In the rest of the article, we will continue to work with the equations (\ref{diffeq0}), (\ref{diffeq1}), (\ref{diffeq2}), (\ref{diffeq3}) and (\ref{diffeq4}) for more generality.

\subsection{Covariance}

Because of their tensorial form, the flexure equations (\ref{diffeqlim0})-(\ref{diffeqlim4}) are covariant; this is also true of the more general equations (\ref{diffeq0}), (\ref{diffeq1}), (\ref{diffeq2}), (\ref{diffeq3}) and (\ref{diffeq4}).
This property means that their form is valid in all systems of coordinates on the sphere, though the tensor components (the covariant derivatives of scalar functions, the metric and the antisymmetric tensor) will have a different expression in each system. The scalar functions will also have a different dependence on the coordinates in each system.
The covariance of the final equations was expected.
We indeed started with tensorial equations in section \ref{3Dtheory}; their restriction to the sphere in principle respected the tensoriality with respect to changes of coordinates on the sphere.
However the covariance of the two-dimensional theory was not made explicit until we obtained the final equations.
This property is thus a strong constraint on the form of the solution and a check of its validity, though only necessary and not sufficient (other covariant terms may have been ignored).

Another advantage of the covariant form is the facility to express the final equations in different systems of coordinates (even non-orthogonal ones) with the aim of solving them.
For example, the finite difference method in spherical coordinates $(\theta,\varphi)$ suffers from a very irregular grid and from pole singularities.
These problems can be avoided with the `cubed sphere' coordinate system \citep{ronchi96}.
Operators including covariant derivatives (here the Laplacian and the operators ${\cal A}$ and ${\cal B}$) can expressed in any system of coordinates whose metric is known.
Christoffel symbols and tensorial differential operators can be computed with symbolic mathematical software.

\subsection{Degree one}
\label{degree1}

Displacements of degree one require special consideration.
All differential operators acting on $(w,F,H)$ (that is $\Delta'$, ${\cal A}$ and ${\cal B}$) in the flexure equations (\ref{diffeq1}) and (\ref{diffeq2}) can be expressed in terms of the operators ${\cal D}_{ij}=\nabla_i \nabla_j + g_{ij}$ (see section \ref{diffop}).
The ${\cal D}_{ij}$ have the interesting property that they give zero when acting on spherical harmonics of degree one (see equation (\ref{propertydeg1})).
Therefore the degree-one terms in the spherical harmonic expansion of $w$ vanish from the flexure equations.
The magnitude of the transverse deflection of degree one neither depends on the load nor on the elastic properties of the spherical shell.

More generally, the homogeneous ($q=\Omega=V=0$) flexure equations (\ref{diffeq0}), (\ref{diffeq1}) and (\ref{diffeq2}) are satisfied by $w$ and $F$ being both of degree one. According to equations (\ref{diffeq3})-(\ref{diffeq4}), the corresponding tangential displacement is constrained by $\Delta S=-2w$ and $\Delta\Delta'T=0$, so that $S$ and $T$ are also of degree one, with $S=w$.
These conditions lead to vanishing strains (see equations (\ref{strainsphere3})) which indicate a rigid displacement.
The total displacement is then given by
\[
{\bf u} = w \, {\bf \hat r} + \bar \nabla w + {\bf \bar \nabla} \times (T{\bf\hat r}) \, ,
\]
where $w$ and $T$ are of degree one.
In Appendix \ref{Adegree1displ}, we show that the first two terms represent a rigid translation whereas the last term represents a rigid rotation.
As expected, stresses vanish for such displacements (see equation (\ref{stresses1})).
This freedom in translating or rotating the solution reflects the freedom in the choice of the reference frame (in practice the reference frame is centered at the center of the undeformed shell).
The same freedom of translation is also found in the theory of deformations of a spherical, radially stratified, gravitating solid \citep[e.g.][]{farrell72,greff97,blewitt03}.

What can we say about degree-one loading?
Let us first examine what happens when the flexure equations are projected on the spherical harmonics of degree one.
Since the operator $\Delta'$ annihilates the degree one in any spherical harmonic expansion, terms of the form $\Delta' f$ vanish when they are projected on the spherical harmonics of degree one.
Moreover, the operators ${\cal A}$ and ${\cal B}$ also vanish in this projection since they do not contain a degree-one term (see equations (\ref{idA})-(\ref{idB}) of Appendix \ref{Adegree1}).
Therefore the degree-one component of the flexure equation (\ref{diffeq2}) is identically zero, whereas the degree-one component of the flexure equation (\ref{diffeq1}) is
\begin{equation}
\int_{\cal S} d\omega \left( q + \bar \nabla \cdot {\bf q}_T \right) Y_i = 0 \hspace{1cm} (i=x,y,z) \, ,
\label{projectionq}
\end{equation}
where $d\omega=\sin \theta \, d\theta \, d\varphi$ and $Y_i$ are the real spherical harmonics of degree one. The integral is taken over the whole spherical surface.
We have used the relation $\Delta\Omega=-R\, \bar \nabla \cdot {\bf q}_T$ derived from equations (\ref{tangentload1}) and (\ref{consoidal}).

The first term in the integrand of equation (\ref{projectionq}) is the projection on the Cartesian axes $({\bf\hat x},{\bf\hat y},{\bf\hat z})$ of the vector field $q\,{\bf\hat r}$:
\[
\left( q \, Y_x , q \, Y_y , q \, Y_z  \right) = \left( q \, {\bf \hat r}  \cdot {\bf \hat x} , q \, {\bf \hat r}  \cdot {\bf \hat y} , q \, {\bf \hat r}  \cdot {\bf \hat z} \right) \, ,
\]
where we have used formulas (\ref{SH1}) for the spherical harmonics.

The second term in equation (\ref{projectionq}) can be rewritten with the identity (\ref{iddivgrad}) and Gauss' theorem (\ref{gauss}):
\[
\int_{\cal S} d\omega \left( \bar \nabla \cdot {\bf q}_T \right) Y_i = - \int_{\cal S} d\omega \;  {\bf q}_T \cdot \bar \nabla \, Y_i  \hspace{1cm} (i=x,y,z) \, ,
\]
where the $Y_i$ are considered as scalars.
Since ${\bf q}_T$ is orthogonal to ${\bf \hat r}$, the integrand is the projection of ${\bf q}_T$ on the Cartesian axes $({\bf\hat x},{\bf\hat y},{\bf\hat z})$:
\[
 \left( {\bf q}_T \cdot \bar \nabla \, Y_x , {\bf q}_T \cdot \bar \nabla \, Y_y , {\bf q}_T \cdot \bar \nabla \, Y_z \right)
 =  \left( {\bf q}_T \cdot {\bf \hat x} , {\bf q}_T \cdot {\bf \hat y} , {\bf q}_T \cdot {\bf \hat z} \right) \, ,
\]
where we have used formulas (\ref{SH1grad}) for the gradients of the spherical harmonics.

Recalling that the transverse load $q$ was defined positive towards the center of the sphere (see equation (\ref{defq})), we define a total load vector ${\bf q}=-q\,{\bf\hat r}+{\bf q}_T$.
With the above results, we can rewrite the degree-one projection (\ref{projectionq}) as
\begin{equation}
\int_{\cal S} d\omega  \left( {\bf q} \cdot {\bf \hat x} \,, {\bf q} \cdot {\bf \hat y} \,, {\bf q} \cdot {\bf \hat z} \right) = ( 0 , 0 , 0 ) \, ,
\end{equation}
which means that the integral (over the whole spherical surface) of the projection on the coordinate axes of the total load vector vanishes.
This result is the consequence of the static assumption in the equations of motion (\ref{stresscurvi}), since a non-zero sum of the external forces would accelerate the sphere.

In practice, degree-one loads on planetary surfaces are essentially due to mass redistribution \citep{greff97} and have a tangential consoidal component (for example the gravitational force is not directed toward the center of figure of the shell).
If the shell thickness is variable, a non-zero $\Omega$ of degree one will induce degrees higher than one in $w$ and in $S$.
If the shell thickness is constant, the degree-one load drops from the flexure equations (\ref{diffeq1})-(\ref{diffeq2}) and $w$ is not affected. However, the degree-one $\Omega$ generates (assuming a constant shell thickness) a degree-one tangential displacement through equation (\ref{diffeq3}), so that $S\neq w$.
Whether the shell thickness is variable or not, a degree-one $\Omega$ thus generates a total displacement which is not only a translation but also a tangential deformation \citep{blewitt03}, in which case stresses do not vanish as shown by equations (\ref{stresses1}).


\section{Limit cases}
\label{limitcases}

\subsection{Membrane limit}
\label{membranelimit}

A shell is in a membrane state of stress if bending moments $(M_\theta,M_\varphi,M_{\theta\varphi})$ can be neglected, in analogy with a membrane which cannot support bending moments.
Equations (\ref{momentres}) show that this is true if the bending rigidity vanishes: $D=0$.
Consistency with equation (\ref{defxi}) imposes the limit of infinite $\xi$ (or $\eta=1$).
With these approximations, the first flexure equation (\ref{diffeq1}) for $(w,F,H)$ becomes
\begin{equation}
\Delta' F = - R \, q - 2 \, \Omega \, .
\label{membrane1}
\end{equation}
The second flexure equation (\ref{diffeq2}) for $(w,F,H)$ becomes
\begin{equation}
\frac{1}{R} \, \Delta' w = \Delta' \, ( \alpha \, \Delta' F ) - (1+\nu) \, {\cal A}(\alpha \,; F) - 2 (1+\nu) \, {\cal B}(\alpha \,; H)
+ (1-\nu) \, \Delta' (\alpha \, \Omega) \, .
\label{membrane2}
\end{equation}
If $q$ is independent of $w$, $F$ and $w$ can be successively determined with spherical harmonic transforms from equations (\ref{membrane1}) and (\ref{membrane2}) (though the right-hand side of the latter equation must be computed with another method).
If $q$ has a linear dependence in $w$ (such as when the sphere is filled with a fluid), $w$ can be eliminated between equations (\ref{membrane1}) and (\ref{membrane2}), so that $F$ and $w$ can also be computed in succession (however the equation for $F$ cannot be solved by a spherical harmonic transform). 
$H$ is supposed to be known since equation (\ref{diffeq0}) is not modified and can be solved with spherical harmonics.

The stresses are obtained from equations (\ref{stresses1}) with the additional approximation of neglecting the term in $\zeta$:
\[
(\hat \sigma_{\theta\theta},\hat \sigma_{\varphi\varphi},\hat \sigma_{\theta\varphi}) =
\frac{1}{h} \left( {\cal O}_2 F + \Omega + 2 {\cal O}_3 H, {\cal O}_1 F + \Omega - 2 {\cal O}_3 H,-{\cal O}_3 F + \left( {\cal O}_2 - {\cal O}_1 \right) H \right) \, .
\]

Bending moments play a small role if the load has a large wavelength.
In practice, the threshold at which bending moments become significant can be evaluated from the constant thickness equation for $w$ (see section \ref{constantthickness}).
One must compare the magnitudes of the terms in $D$ and $1/\alpha$ in the left-hand side of equation (\ref{diffeqw}).
Bending moments are negligible (i.e. the term in $D$) if the spherical harmonic degree $\ell$ of the transverse displacement $w$ is such that
\begin{equation}
\ell \lesssim  k \, \sqrt{ \frac{R}{h} } \, ,
\label{membranecond}
\end{equation}
where $k=(12(1-\nu^2))^{1/4}\approx1.8$ if we take $\nu=1/4$.
This threshold is about 10 for a planet with a radius of 3400 km and a lithospheric thickness equal to 100 km.
Flexure equations in the membrane limit of a shell with constant thickness have been used by \citet{sleep85} to study the lithospheric stress in the Tharsis region of the planet Mars.

\subsection{Euclidean limit}
\label{euclideanlimit}

The equation for the deflection of a rectangular plate with variable thickness has been derived by \citet[p.~173]{timoshenkoW}.
We will check here that the Euclidean limit of our equations gives the same answer.

Let us define the coordinates $(x,y)=(R\,\varphi,R\,\theta')$, where $\theta'=\theta/2-\theta$ is the latitude.
We work in a small latitude band around the equator (so that $\theta'\ll1$) and in the limit of large spherical radius $R$ and large $\xi$ ($\eta=1$).
Under this change of coordinate, each derivative introduces a factor $R$ so that terms with the largest number of derivatives dominate.
In particular, covariant derivatives can be approximated by usual derivatives.
The surface Laplacian (\ref{laplacian}) can be approximated as follows:
\begin{eqnarray}
\Delta
&\approx& R^2 \left( \frac{\partial^2}{\partial x^2} + \frac{\partial^2}{\partial y^2} \right) \, ,
\nonumber \\
&\equiv& R^2 \, \Delta_e \, .
\nonumber
\end{eqnarray}
We assume that there are no tangential loads ($\Omega=V=H=0$).
The flexure equations (\ref{diffeqlim1})-(\ref{diffeqlim2}) for the transverse displacement become:
\begin{eqnarray}
R^4 \, \Delta_e  \left( D \, \Delta_e w \right)
- (1-\nu) \, R^4 \, {\cal A}_e (D \,; w)
+ R^5 \Delta_e F &=& -R^4 \, q \, ,
\label{eucli1} \\
R^4 \, \Delta_e \left( \, \alpha \, \Delta_e F \right)
- (1+\nu) \, R^4 \, {\cal A}_e (\alpha \,; F)
- R \; \Delta_e w &=& 0 \, ,
\label{eucli2}
\end{eqnarray}
where the operator ${\cal A}_e$ is defined by
\begin{eqnarray}
{\cal A}_e (a \,; b)
&=& \left( \Delta_e \, a \right) \left( \Delta_e \, b \right) - a_{, i , j} \, b_{, i , j}
\nonumber \\
&=& \left( \frac{\partial^2 a}{\partial x^2} \right) \left( \frac{\partial^2 b}{\partial y^2} \right)
- 2 \, \left( \frac{\partial^2 a}{\partial x \partial y} \right) \left(\frac{\partial^2 b}{\partial x \partial y} \right)
+ \left(\frac{\partial^2 a}{\partial y^2} \right) \left(\frac{\partial^2 b}{\partial x^2} \right) \, .
\nonumber
\end{eqnarray}
Equation (\ref{eucli2}) gives a relation between the magnitudes of $F$ and $w$:
\[
{\cal O} ( R^3 \, \Delta_e F ) \sim {\cal O}(w/\alpha) \, .
\]
In the large $R$ limit, equation (\ref{eucli1}) thus becomes
\[
\Delta_e  \left( D \, \Delta_e w \right) - (1-\nu) \, {\cal A}_e (D \,; w) = - q \, ,
\]
which is the equation derived by \citet{timoshenkoW}.
This equation has been used by \citet{stark03}, \citet{kirby04} and \citet{perezgussinye04} for the local analysis of the lithosphere of the Earth.
Its one-dimensional version, in which ${\cal A}_e$ vanishes, has been used by \citet{sandwell84} to describe the flexure of the oceanic lithosphere on Earth and by \citet{stewart97} to model the flexure at mountain ranges.

\subsection{Shell with constant thickness}

\subsubsection{Displacements}
\label{constantthickness}

If the thickness of the shell is constant, the toroidal part of the tangential displacement decouples.
The terms in ${\cal B}$ indeed drop from the flexure equations (\ref{diffeq1}) and (\ref{diffeq2}) so that the equations for $(w,F)$ depend only on $(q,\Omega)$:
\begin{eqnarray}
\eta D \, \Delta' \Delta' w - (1-\nu) \, \eta D \, \Delta' w + \eta R^3 \, \Delta' F
&=& -R^4 \, q + R^3 \left( (1-\eta) \, \Delta \Omega - 2 \eta \Omega \right) \, ,
\label{diffeqcst1} \\
\Delta' \Delta' F - (1+\nu) \,  \Delta' F - \frac{1}{R\alpha} \, \Delta' w 
&=& - (1-\nu) \, \Delta' \Omega \, ,
\label{diffeqcst2}
\end{eqnarray}
where we used the property ${\cal A}(a\,;b)=a\Delta'b$ valid for constant $a$.
We eliminate $F$ from these equations and obtain a sixth order equation relating $w$ to $(q,\Omega)$:
\begin{equation}
\eta D \, \Delta \, \Delta' \Delta' w + \frac{R^2}{\alpha} \, \Delta' w = - R^4 \left( \Delta' -1-\nu \right) q + R^3 \left( \frac{1}{1+\xi} \, \Delta' -1-\nu \right) \Delta \Omega \, .
\label{diffeqw}
\end{equation}

The elimination of $\Delta'F$ between equations (\ref{diffeqlim3}) and (\ref{diffeqcst1}) gives an equation relating the consoidal displacement potential $S$ to $(w,q,\Omega)$:
\[
\Delta \, S = - \frac{1}{1+\nu} \, \frac{1}{1+\xi} \, \Delta \, \Delta' w  - 2 \, w - (1-\nu) \, R^2 \alpha \, q + \frac{1-\nu}{1+\xi} \, R \alpha \, \Delta \Omega \, .
\]
Terms not including a Laplacian can be eliminated with equation (\ref{diffeqw}), so that we obtain an explicit solution for $S$ in terms of $(w,q,\Omega)$:
\begin{equation}
S = \frac{1}{1+\xi} \, \frac{1}{1-\nu^2} \left( \Delta + 1 + \nu \right) \Delta' w + w + R^2 \alpha \, q - \frac{R \alpha}{1+\xi} \left( \Delta - \xi (1+\nu) \right) \Omega \, ,
\label{solS} 
\end{equation}
where the integration constant has been set to zero.

Equations (\ref{diffeq0}) and (\ref{diffeq4}) give an equation for the toroidal displacement potential:
\begin{equation}
\Delta' \, T = - 2 \eta R \alpha \, (1+\nu) \, V \, ,
\label{diffeqT}
\end{equation}
where the integration constant has been set to zero. We have assumed that $\Delta V\neq0$, otherwise we get $\Delta'T=0$.

The differential equations given in this section can be solved with spherical harmonics so that the coefficients of the spherical harmonic expansions of $(w,S,T)$ can be expressed in terms of the corresponding coefficients of the loads $(q,\Omega,V)$ (see \citet{kraus}, \citet{turcotte81}, \citet{banerdt86}).

\subsubsection{Comparison with the literature}

We now compare our equations for a shell of constant thickness with those found in the literature. 
The formulas of \citet{banerdt86} (taken from the work of \citet{vlasov}) are the most general:
\begin{eqnarray}
D \left( \Delta^3 + 4 \Delta^2 \right) w + \frac{R^2}{\alpha} \left( \Delta + 2 \right) w
&=& - R^4 \left( \Delta + 1-\nu \right) q + R^3 \left( \frac{1}{\xi} \, \Delta -1-\nu \right) \Delta \Omega \, ,
\label{eqbanerdt1}
\\
\left( \Delta +2 \right) \chi &=& \frac{R^2}{D\xi(1-\nu)} \, \Delta \, V \, .
\label{eqbanerdt2}
\end{eqnarray}
Banerdt's notation is slightly different: his formulas are obtained with the substitutions $\xi\rightarrow\psi$, $\Omega\rightarrow\,R\Omega$ and $V\rightarrow RV$.
The {\it normal rotation} $\chi$ is proportional to the radial component (in a normalized basis) of the curl of the tangential displacement:
\[
\chi = \frac{1}{2R} \left( \nabla \times {\bf v} \right)_{\hat r} \, .
\]
The curl $\nabla\times{\bf v}$ is related to our surface curl (\ref{surfacecurl}) by
\[
\nabla \times {\bf v} = \bar \nabla \times {\bf v} + \csc \theta \left( \left( \sin \theta \hat v_\varphi \right)_{, \theta} - \hat v_{\theta , \varphi} \right) {\bf \hat r} \, .
\]
With the formulas (\ref{tangentdispl2}) and (\ref{laplacian}), we get $\nabla\times{\bf v}=-\Delta \,T\,{\bf \hat r}$ so that equation (\ref{eqbanerdt2}) becomes
\begin{equation}
\Delta' \, T = - 2 R \alpha (1+\nu) \, V \, .
\label{eqbanerdt2bis}
\end{equation}
We see that Banerdt's equations (\ref{eqbanerdt1}) and (\ref{eqbanerdt2bis}) coincide with our equations (\ref{diffeqw}) and (\ref{diffeqT}) in the limit of large $\xi$ ($\eta=1$), with one exception: the bending term for $w$ is written $D(\Delta^3+4\Delta^2)w$ instead of $D\Delta\,\Delta'\Delta'w=(\Delta^3+4\Delta^2+4\Delta)w$.
This error has propagated in many articles and is of consequence for the degree-one harmonic component, since it violates the static assumption and spoils the translation invariance discussed in section \ref{degree1}. The impact on higher degrees is negligible.
Because of this mistake, many authors give a separate treatment to the first harmonic degree.
Banerdt also gives formulas for the tangential displacements in terms of consoidal and toroidal scalars $(A,B)$ corresponding to our scalars $(S,T)$: his formula (A10) is equivalent to our equation (\ref{solS}) in the limit of large $\xi$.

If we ignore temperature effects, Kraus' first equation for $(w,F)$ is equivalent to our equation (\ref{diffeqcst1}) in the limit of large $\xi$, whereas his second equation for $(w,F)$ is equivalent to the combination eq.(\ref{diffeqcst2})$\,+\,\frac{1+\nu}{\eta R^3}\,$eq.(\ref{diffeqcst1}) in the limit of large $\xi$ \citep[see][eq. 6.54h and 6.55d]{kraus}.
Note that the definition of Kraus' stress function $F$ \citep[p.~243]{kraus} differs from ours:
\[
\frac{1}{R^2} \, F_{Kraus} = F - k (1-\nu) \, \frac{D}{R^3} \, w \, ,
\]
with $k=1$.
This freedom of redefining $F$ for arbitrary $k$ remains as long as $D$ is constant. The flexure equation for $w$, equation (\ref{diffeqw}), is unaffected so that the solution for $w$ is unchanged.
In the final step, Kraus makes a mistake when combining the two equations for $(w,F)$ and thus obtains a flexure equation for $w$ with the same error as in equation (\ref{eqbanerdt1}). Kraus does not include toroidal loading.
The flexure equation of \citet{turcotte81} is taken from \citet{kraus} without the tangential loading and is the same as equation (\ref{eqbanerdt1}) with $\Omega=0$.

The flexure equation of \citet{brotchie69} is given in our notation by
\begin{equation}
D \, \Delta^2 w + \frac{R^2}{\alpha} \, w = - R^4 \, q \, ,
\label{eqbrotchie}
\end{equation}
where $q$ includes their term $\gamma w$ describing the response of the enclosed liquid.
This equation can be obtained from our equation (\ref{diffeqw}) as follows: keep only the derivatives of the highest order in each term, set $\Omega=0$, take the limit of large $\xi$ ($\eta=1$) and integrate.
Brotchie and Silvester choose to work in the approximation of a shallow shell and with axisymmetrical loading, solving their equation in polar coordinates with Bessel-Kelvin functions.
The reduction to fourth order in equation (\ref{eqbrotchie}), the shallow shell approximation and the axisymmetrical assumption are not justified nowadays since the full equation (\ref{diffeqw}) can be quickly solved with computer-generated spherical harmonics.

The contraction due to a transverse load of degree~0, $w=-R^2\alpha(1-\nu)q/2$, is equivalent to the radial displacement computed by Love in the limit of a thin shell \citep[][p.142]{love}. However additional assumptions about the initial state of stress and the internal density changes are necessary \citep{willemann82} so that the degree~0 is usually excluded from the analysis.

\subsubsection{Breakdown of the third assumption of thin shell theory}
\label{breakdown}

With the spherical harmonic solutions of the equations for a shell of constant thickness, it is possible to check the thin shell assumption stating that the transverse normal stress is negligible with respect to the tangential normal stress. The magnitude of the former can be estimated by the load $q$ (see definition (\ref{defq})) whereas the magnitude of the latter can be approximated with formulas (\ref{stresses1}) evaluated on the outer surface:
\begin{eqnarray}
\sigma_T
&\equiv& \frac{1}{2} \left( \hat \sigma_{\theta\theta} + \hat \sigma_{\varphi\varphi} \right)|_{\frac{h}{2}}
\nonumber \\
&\approx& \frac{1}{2h} \, \Delta' F - \frac{Eh}{4R^2(1-\nu)} \, \Delta' w \, .
\nonumber
\end{eqnarray}
where we have assumed the absence of tangential loads ($\Omega=0$) and the limit of large $\xi$.
We can relate $\sigma_T$ to $q$ by using the solution in spherical harmonics of equations (\ref{diffeqcst2}) and (\ref{diffeqw}).
Since the thin shell assumption is expected to fail for a load of sufficiently small wavelength, we assume that the spherical harmonic degree $\ell$ is large.
Assuming $\ell\gg1$, we obtain
\begin{eqnarray}
(\Delta' F)_{\ell m} &\approx& \frac{1}{R\alpha} \, w_{\ell m} \, ,
\nonumber \\
w_{\ell m} &\approx& - \frac{R^2\alpha}{1+\frac{\ell^4}{\xi (1-\nu^2) }} \, q_{\ell m} \, ,
\nonumber
\end{eqnarray}
where the spherical harmonic coefficients are indexed by their degree $\ell$ and their order $m$.
If the shell is not in a membrane state of stress (see equation (\ref{membranecond})), $\ell^2>2R/h$ so that $\sigma_T$ can be approximated by
\[
(\sigma_T)_{lm} \approx \frac{\xi (1+\nu) }{4\ell^2} \, q_{lm} \, .
\]
The thin shell assumption holds if $q<\sigma_T$, that is if
\begin{equation}
\ell <  \sqrt{3 (1+\nu) } \, \frac{R}{h}
\hspace{1cm}
\mbox{or}
\hspace{1cm}
\lambda > \frac{2 \pi}{\sqrt{3(1+\nu)}} \, h \, ,
\label{thinshellcond1}
\end{equation}
where $\lambda$ is the load wavelength ($\lambda\approx2\pi R/\ell$).
We have $\sqrt{3 (1+\nu)}\approx1.9$  and $2\pi/\sqrt{3(1+\nu)}\approx3.2$ if we take $\nu=1/4$.
This condition on $\lambda$ is consistent with the transition zone between the thin and thick shell responses analyzed in \citet{janes90} and \citet{zhong00},
but does not coincide with the constraint given in \citet{willemann82}, which is $\ell<2\pi\sqrt{R/h}$ (this last condition looks more like the threshold (\ref{membranecond}) for the membrane regime).

Though the stress distribution is affected, the limit (\ref{thinshellcond1}) on the degree $\ell$ is not important for the displacements, since they tend to zero at small wavelengths.
Therefore the theory does not break down at short wavelength if one is interested in the computation of the gravity field associated to the transverse deflection of the lithosphere.


\section{Conclusion}

The principal results of this article are the five flexure equations (\ref{diffeqlim0})-(\ref{diffeqlim4}) governing the three displacements of the thin spherical shell and the two auxiliary stress functions.
Stresses are derived quantities which can be obtained from equations (\ref{stresses1}).
The shell thickness and Young's modulus can vary, but Poisson's ratio must be constant.
The loads acting on the shell can be of any type since we extend the method of stress functions to include not only transverse and consoidal tangential loads, but also toroidal tangential loads.
The flexure equations can be solved one after the other, except the two equations (\ref{diffeqlim1})-(\ref{diffeqlim2}) for the transverse deflection $w$ and the stress function $F$, which must be simultaneously solved.
Tangential loading is usually neglected when solving for the deflection because of its small effect.
In that case, it is sufficient to solve the two equations (\ref{diffeqlim1})-(\ref{diffeqlim2}) with $\Omega=H=0$:
\begin{eqnarray}
\Delta'  \left( D \, \Delta' w \right) - (1-\nu) \, {\cal A}(D \,; w) + R^3 \Delta' F &=& -R^4 \, q\, ,
\nonumber \\
\Delta' \left( \alpha \, \Delta' F \right) - (1+\nu) \, {\cal A}( \alpha \,; F) - \frac{1}{R} \; \Delta' w &=& 0 \, .
\nonumber
\end{eqnarray}
However tangential loading must be taken into account when computing stress fields \citep{banerdt86}.

In the long-wavelength limit (i.e. membrane regime), all equations can be solved one after the other because it is possible to solve for $F$ before solving for the transverse deflection.
If a small part of the shell is considered, the flexure equations reduce to the equations governing the deflection of a flat plate with variable thickness.
If the shell thickness is constant, the flexure equations reduce to equations available in the literature which can be completely solved with spherical harmonics.
Our rigorous treatment of the thin shell approximation has clarified the effect of the shell thickness on the flexure equations.
We emphasize the need to use the correct form for the equations (without the common mistake in the differential operator acting on $w$) in order to have the correct properties for the degree-one deflection and degree-one load.

We have also obtained two general properties of the flexure equations.
First we have shown that there is always a toroidal component in the tangential displacement if the shell thickness is variable.
Second we have proven that the degree-one harmonic components of the transverse deflection and of the toroidal component of the tangential displacement do not depend on the elastic properties of the shell. This property reflects the freedom under translations and rotations of the reference frame. Besides we have shown that degree-one loads are constrained by the static assumption but can deform the shell and generate stresses.

This article was dedicated to the theoretical treatment of the flexure of a thin elastic shell with variable thickness.
While the special case of constant thickness admits an analytical solution in terms of spherical harmonics, the general flexure equations must be solved with numerical methods such as finite differences, finite elements or pseudospectral methods.
In a forthcoming paper, we will give a practical method of solution and discuss applications to real cases.

\subsection*{Acknowledgments}
M. Beuthe is supported by a PRODEX grant of the Belgian Science Federal Policy.
The author thanks Tim Van Hoolst for his help and Jeanne De Jaegher for useful comments.
Special thanks are due to Patrick Wu for his constructive criticisms which helped to improve the manuscript.


\section{Appendix}

\subsection{Covariant, contravariant and normalized components}
\label{Acomponents}

Tensors can be defined by their transformation law under changes of coordinates.
The two types of tensor components, namely {\it covariant} and {\it contravariant} components, transform in a reciprocal way under changes of coordinates.
Tensor components cannot be expressed in a normalized basis: the space must have a coordinate vector basis (for contravariant components) and a dual basis (for covariant components) which are not normalized.

The only exception is a flat space with Cartesian coordinates, where covariant, contravariant and normalized components are identical.
Since the metric is the scalar product of the elements of the coordinate vector basis, the covariant components are related to components defined in a normalized basis (written with a hat) by
\[
u_i = \sqrt{g_{ii}} \; \hat u_i \, ,
\]
whereas the relation for contravariant components is
\[
u^i = \frac{1}{\sqrt{g_{ii}}} \; \hat u_i \, .
\]

The normalized Cartesian basis $({\bf\hat x},{\bf\hat y},{\bf\hat z})$ is related to the normalized basis for spherical coordinates $({\bf\hat r},\pb,\tb)$ by
\begin{eqnarray}
{\bf\hat x} &=& \cos \theta \cos \varphi \, \pb - \sin \varphi \, \tb + \sin \theta \cos \varphi \, {\bf \hat r} \, ,
\nonumber \\
{\bf\hat y} &=& \cos \theta \sin \varphi \, \pb + \cos \varphi \, \tb + \sin \theta \sin \varphi \, {\bf \hat r} \, ,
\label{changebasis} \\
{\bf\hat z} &=& - \sin \theta \, \pb + \cos \theta \, {\bf \hat r} \, .
\nonumber
\end{eqnarray}

\subsection{Covariant derivatives}
\label{Aderivatives}

Usual derivatives are indicated by a `comma':
\[
v_{i,j}=\frac{\partial v_i}{\partial x^j} \, .
\]
Covariant derivatives (defined below) are indicated by a `bar' or by the operator $\nabla_i$:
\[
v_{i | j} = \nabla_j \, v_i \, .
\]
The former notation emphasizes the tensorial character of the covariant derivative since the covariant derivative adds a covariant index to the vector.
The latter notation is more adapted when we are interested by the properties of the operator.

The covariant derivative of a scalar function $f$ is equal to the usual derivative, $f_{| i} = f_{, i}$, and is itself a covariant vector: $f_{| i}=v_i$.
Covariant derivatives on covariant and contravariant vector components are defined by
\begin{eqnarray}
v_{i | j} &=& v_{i,j} - \Gamma_{ij}^k \, v_k \, ,
\label{covariant} \\
v^i_{\;\; | j} &=& v^i_{\;\; , j} + \Gamma_{jk}^i \, v^k \, ,
\label{contravariant}
\end{eqnarray}
where the summation on repeated indices is implicit.
The symbols $\Gamma_{ij}^k$ are the Christoffel symbols of the second kind \citep{synge}.
Their expressions for the metrics used in this article are given in sections \ref{A3Dgeom} and \ref{A2Dgeom}.

Covariant differentiation of higher order tensors is explained in \citet{synge} but we only need the rule for a covariant tensor of second order:
\[
\sigma_{ij | k} = \sigma_{ij , k} - \Gamma_{ik}^l \, \sigma_{lj} - \Gamma_{jk}^l \, \sigma_{il} \, .
\]
If some of the indices of the tensor are contravariant, the rule is changed according to equation (\ref{contravariant}). 
The covariant derivatives of the metric and of the inverse metric are zero: $g_{ij | k}=0$ and $g^{ij}_{\;\;\; | k}=0$.

\subsection{Three-dimensional spherical geometry}
\label{A3Dgeom}

The geometry of a thin spherical shell of average radius $R$ can be described with coordinates $\theta$, $\varphi$ and $\zeta$, respectively representing the colatitude, longitude and radial coordinates. The radial coordinate $\zeta$ is zero on the reference surface (i.e. the sphere of radius $R$) of the shell. The non-zero components of the metric are given by
\begin{eqnarray}
g_{\theta\theta} &=& (R+\zeta)^2 \, ,
\nonumber \\
g_{\varphi\varphi} &=& (R+\zeta)^2 \, \sin^2 \theta \, ,
\nonumber \\
g_{\zeta\zeta} &=& 1 \, .
\nonumber
\end{eqnarray}
The non-zero Christoffel symbols are given by
\begin{eqnarray}
\Gamma_{\theta\theta}^\zeta &=& -(R+\zeta) \, ,
\nonumber \\
\Gamma_{\varphi\varphi}^\zeta &=& -(R+\zeta) \, \sin^2 \theta \, ,
\nonumber \\
\Gamma_{\zeta\theta}^\theta &=& \Gamma_{\theta\zeta}^\theta \; = \; \Gamma_{\zeta\varphi}^\varphi \; = \; \Gamma_{\varphi\zeta}^\varphi \; = \; \frac{1}{R+\zeta} \, ,
\nonumber \\
\Gamma_{\varphi\varphi}^\theta &=& - \sin \theta \, \cos \theta \, ,
\nonumber \\
\Gamma_{\varphi\theta}^\varphi &=& \Gamma_{\theta\varphi}^\varphi = \cot \theta \, .
\nonumber
\end{eqnarray}

\subsection{Two-dimensional spherical geometry}
\label{A2Dgeom}

If $\theta$ and $\varphi$ respectively represent the colatitude and longitude coordinates, the non-zero components of the metric on the surface of the sphere are given by
\begin{eqnarray}
g_{\theta\theta} &=& 1 \, ,
\nonumber \\
g_{\varphi\varphi} &=& \sin^2 \theta \, .
\label{metric2D}
\end{eqnarray}
The non-zero Christoffel symbols are given by
\begin{eqnarray}
\Gamma_{\varphi\varphi}^\theta &=& - \sin \theta \, \cos \theta \, ,
\nonumber \\
\Gamma_{\varphi\theta}^\varphi &=& \Gamma_{\theta\varphi}^\varphi \; = \; \cot \theta \, .
\nonumber
\end{eqnarray}
The double covariant derivatives of a scalar function $f$ are thus given by
\begin{eqnarray}
f_{| \theta | \theta} &=& f_{, \theta , \theta} \, ,
\nonumber \\
f_{| \theta | \varphi} &=& f_{| \varphi | \theta} = f_{, \theta , \varphi} - \cot \theta \, f_{, \varphi} \, ,
\nonumber \\
f_{| \varphi | \varphi} &=& f_{, \varphi , \varphi} + \sin \theta \,  \cos \theta \, f_{, \theta} \, .
\nonumber
\end{eqnarray}

An antisymmetric tensor $\varepsilon_{ij}$ is defined by
\[
\varepsilon_{ij} \equiv \sqrt{ \det g_{ij} } \; \bar \varepsilon_{ij} \, ,
\]
where $\bar \varepsilon_{ij}$ is the antisymmetric symbol invariant under coordinate transformations:
$\bar \varepsilon_{\theta\varphi}=-\bar \varepsilon_{\varphi\theta}=1$,
$\bar \varepsilon_{\theta\theta}=\bar \varepsilon_{\varphi\varphi}=0$
($\bar \varepsilon_{ij}$ is usually called a tensor density; \citet{synge} call it a relative tensor of weight -1).
The non-zero covariant components of $\varepsilon_{ij}$ are given for the metric of the spherical surface by
\[
\varepsilon_{\theta\varphi} = - \varepsilon_{\varphi\theta} = \sin \theta \, .
\]
The non-zero contravariant components, $\varepsilon^{ij}=g^{ik}g^{jl}\varepsilon_{kl}$, are given by
\[
\varepsilon^{\theta\varphi} = - \varepsilon^{\varphi\theta} = \, \csc \theta \, .
\]
The covariant derivative of the tensor $\varepsilon_{ij}$ is zero: $\varepsilon_{ij|k}=0$.

\subsection{Gradient, divergence, curl and Laplacian}
\label{Adiffop}

Various differential operators on the surface of the sphere can be constructed with covariant derivatives.
In this section, $f$ and $t$ are scalar functions defined on the sphere and ${\bf v}$ is a vector tangent to the sphere.
\citet{backus86} gives more details on surface operators and on Helmholtz's theorem.

As mentioned in Appendix \ref{Aderivatives}, the covariant derivative of a scalar function $f$ defined on the sphere is a covariant vector tangent to the sphere whose components are $f_{,\theta}$ and $f_{,\varphi}$.
The {\it surface gradient} of $f$ is the same vector with its components expressed in the normalized basis $(\pb,\tb)$:
\begin{equation}
\bar \nabla f = f_{,\theta} \, \pb + \csc \theta \, f_{,\varphi} \, \tb \, .
\label{surfacegrad}
\end{equation}

The contraction of the covariant derivative with the components of a vector ${\bf v}$ yields a scalar:
\[
v^i_{\;\; | i} = v^\theta_{\;\; , \theta} + \cot \theta \, v^\theta + v^\varphi_{\;\; , \varphi} \, .
\]
The {\it surface divergence} is the corresponding operation on the vector with its components expressed in the normalized basis $(\pb,\tb)$:
\begin{equation}
\bar \nabla \cdot {\bf v} = \csc \theta \left( \left( \sin \theta \, \hat v_\theta \right)_{, \theta}  + \hat v_{\varphi , \varphi} \right) \, .
\label{surfacediv}
\end{equation}
Since the result is a scalar, $v^i_{\;\; | i}=\bar\nabla\cdot{\bf v}$.
A useful identity is
\begin{equation}
\bar \nabla \cdot \left( f \, {\bf v} \right) = \bar \nabla f \cdot {\bf v} + f \, \bar \nabla \cdot {\bf v} \, .
\label{iddivgrad}
\end{equation}

The contraction of the antisymmetric tensor $\varepsilon_{ij}$ with the covariant derivative of a scalar $t$ yields the covariant components of a vector ${\bf v}$:
\[
v_i = g^{jk} \, \varepsilon_{ik \, } t_{, j} \, .
\]
The components are given for the metric (\ref{metric2D}) by $v_\theta=\csc\theta\,t_{,\varphi}$ and $v_\varphi=-\sin\theta\,t_{,\theta}$.
If $t$ is considered as the radial component of the radial vector ${\bf t}=t\,{\bf\hat r}$ (the covariant radial component is equal to the normalized one), $v_i$ are the non-zero covariant components of the three-dimensional curl of ${\bf t}$, which is tangent to the sphere.
This fact justifies the definition of the {\it surface curl} of ${\bf t}$, which is equal to the vector ${\bf v}$ but with components given in the normalized basis $(\pb,\tb)$:
\begin{equation}
\bar \nabla \times {\bf t} = \csc \theta \, t_{, \varphi} \, \pb - t_{, \theta} \, \tb \, .
\label{surfacecurl}
\end{equation}

The contraction of the double covariant derivative acting on a scalar $f$ defines the {\it surface Laplacian}:
\begin{eqnarray}
\Delta f
&=& g^{ij} \, f_{| i | j}
\nonumber \\
&=& f_{, \theta , \theta} + \cot \theta \, f_{, \theta} + \csc^2 \theta \, f_{, \varphi , \varphi} \, .
\label{laplacian}
\end{eqnarray}
The surface Laplacian can also be seen as the composition of the surface divergence with the surface gradient: $\Delta f=\bar\nabla\cdot\bar\nabla f$.

According to Helmholtz's theorem, a vector tangent to the sphere can be written as the sum of the surface gradient of a scalar $f$ and the surface curl of a radial vector $t \, {\bf \hat r}$:
\begin{equation}
{\bf v} = \bar \nabla f + \bar \nabla \times ( t \, {\bf \hat r}) \, .
\label{helmholtz}
\end{equation}
While $t$ is always called the {\it toroidal} scalar (or potential) for ${\bf v}$, there is no standard terminology for $f$.
\citet{backus86} calls $f$ the {\it consoidal} scalar for ${\bf v}$.
Some authors \citep[e.g.][]{banerdt86} call $f$ the {\it poloidal} potential for ${\bf v}$. The origin of this use lies in the theory of mantle convection, in which plate tectonics are assumed to be driven by mantle flow. Under the assumption of an incompressible mantle fluid, the velocity field of the fluid is solenoidal, i.e. its 3-dimensional divergence vanishes.
In such a case, the velocity field can be decomposed into a poloidal part ($\nabla\times\nabla\times(P\,{\bf \hat r})$) and a toroidal part ($\nabla\times(Q\,{\bf \hat r})$), where differential operators are 3-dimensional \citep{backus86}.
If the velocity field is tangent to the spherical surface, the poloidal component at the surface is also the consoidal component \citep{forte87}.
However the fields for which we use Helmholtz's theorem, i.e. the tangential surface load and the tangential surface displacement, do not belong to 3-dimensional solenoidal vector fields. We thus prefer to use the term `consoidal'.

The surface divergence of ${\bf v}$ depends only on the consoidal scalar $f$:
\begin{equation}
\bar \nabla \cdot {\bf v} = \Delta f \, .
\label{consoidal}
\end{equation}

The two-dimensional version of Gauss theorem is
\begin{equation}
\int_{\cal S} d\omega \; \bar \nabla \cdot {\bf v} = 0 \, .
\label{gauss}
\end{equation}
where $d\omega=\sin \theta \, d\theta \, d\varphi$ and the integral is taken over the whole spherical surface.
It can be proven with formula (\ref{surfacediv}).

\subsection{Rigid displacements}
\label{Adegree1displ}

At the surface of a sphere subjected to deformation, the displacement ${\bf u}$ of a point can be expressed with the help of Helmholtz's theorem (\ref{helmholtz}) in terms of three scalar functions $(w,S,T)$ depending on $\theta$ and $\varphi$:
\begin{equation}
  {\bf u} = w \, {\bf \hat r} + \bar \nabla S + {\bf \bar \nabla} \times (T{\bf\hat r}) \, .
\end{equation}

Strains (and stresses) vanish for rigid displacements. Equations (\ref{strainsphere3}) show that strains vanish when $(w,S,T)$ are of degree one, with $S=w$ (recall that the operators ${\cal O}_i$ annihilate the degree one).
Assuming these conditions, we now show that ${\bf u}_{transl}=w\,{\bf\hat r}+\bar\nabla\,w$ represents a rigid translation whereas ${\bf u}_{rot}={\bf \bar \nabla} \times (t\,{\bf\hat r})$ represents a rigid rotation of the sphere.
We choose as basis the real spherical harmonics of degree one which form the components of the radial unit vector in Cartesian coordinates: 
\begin{eqnarray}
(Y_x,Y_y,Y_z)
&=& (\sin\theta\cos\varphi,\sin\theta\sin\varphi,\cos\theta)
\nonumber \\
&=& \left( {\bf\hat x},{\bf\hat y},{\bf\hat z} \right) \cdot {\bf \hat r} \, .
\label{SH1}
\end{eqnarray}
We need the surface gradient of the real spherical harmonics which can be computed with formulas (\ref{changebasis}) and (\ref{surfacegrad}):
\begin{equation}
\left(
\bar \nabla \, Y_x,
\bar \nabla \, Y_y,
\bar \nabla \, Y_z
\right)
=
\left(
{\bf \hat x} - \sin \theta \cos \varphi \, {\bf \hat r},
{\bf \hat y} - \sin \theta \sin \varphi \, {\bf \hat r},
{\bf \hat z} - \cos \theta \, {\bf \hat r}
\right) \, .
\label{SH1grad}
\end{equation}

If the expansion of $w$ in the degree-one basis is $w=a\,Y_x+b\,Y_y+c\,Y_z$, then
\[
 w \, {\bf \hat r} + \bar \nabla\,w = a \, {\bf \hat x} + b \, {\bf \hat y} + c \, {\bf \hat z} \, ,
\]
so that ${\bf u}_{transl}$ is indeed a rigid translation of the sphere.

If the expansion of $T$ in the degree-one basis is $T=a'Y_x+b'Y_y+c'Y_z$, then
\[
{\bf \bar \nabla} \times (T\,{\bf\hat r}) =
a' \left( -\sin \varphi \, \pb - \cos \theta \cos \varphi \, \tb \right)
+ b' \left( \cos \varphi \, \pb - \cos \theta \sin \varphi \, \tb \right)
+ c' \sin \theta \, \tb \, ,
\]
so that ${\bf u}_{rot}$ includes a rigid rotation of the sphere, with $(a',b',c')$ being the angles of rotation around the axes $({\bf \hat x},{\bf \hat y},{\bf \hat z})$, respectively.
Though ${\bf u}_{rot}$ seems to include a uniform radial expansion, one should recall that linearized strain-displacement equations are not valid for large displacements.
Since strains vanish, the radial expansion is not physical and ${\bf u}_{rot}$ represents a pure rotation.
Finite deformations are for example discussed in \citet{love}[pp.~66-73] and \citet{sokolnikoff}[pp.~29-33].

\subsection{Differential identities for the operators ${\cal O}_i$}
\label{Aidentities}

The differential operators ${\cal O}_i$ defined by equations (\ref{opO}) satisfy differential identities useful when obtaining the flexure equations.
They are special cases of differential identities valid in curved spaces.
The presence of curvature makes the parallel transport of vectors path-dependent;
this property quantifies the curvature of space and can be expressed as the lack of commutativity of the covariant derivatives of a vector ${\bf v}$:
\begin{equation}
v_{i | j | k} - v_{i | k | j} = R_{iljk} \, v^l \, ,
\label{commut}
\end{equation}
where $R_{iljk}$ are the covariant components of the Riemann tensor.
On the sphere, the Riemann tensor has only one independent component that is non-zero,
$R_{\theta\varphi\theta\varphi} = - \sin^2 \theta$.
Other components are related by the symmetries
$R_{\alpha\beta\gamma\delta}=-R_{\beta\alpha\gamma\delta}=-R_{\alpha\beta\delta\gamma}=R_{\gamma\delta\alpha\beta}$.

The substitution of $f_{, i}$ to $v_i$ in the commutation relation (\ref{commut}) provides two differential identities satisfied by double covariant derivatives acting on scalar functions:
\begin{eqnarray}
\left( \csc \theta \, f_{| \varphi | \varphi} \right)_{, \theta} - \csc \theta \, f_{| \varphi | \theta , \varphi }  - \cos \theta \, f_{| \theta | \theta} + \sin \theta \, f_{, \theta} &=& 0 \, ,
\nonumber \\
f_{| \theta | \theta , \varphi} - f_{| \varphi | \theta , \theta} - \cot \theta \, f_{| \varphi | \theta} + f_{, \varphi} &=& 0 \, .
\nonumber
\end{eqnarray}

The replacement  in the above equations of the double covariant derivatives by the normalized differential operators (\ref{opObis}) yields the following identities:
\begin{eqnarray}
\left( \sin \theta \, {\cal O}_2 f \right)_{, \theta} - \left( {\cal O}_3 f \right)_{, \varphi} - \cos \theta \, {\cal O}_1 f &=& 0 \hspace{1cm} \mbox{(I1)} \, ,
\label{id1} \\
\left( \sin \theta \, {\cal O}_3 f \right)_{, \theta} - \left( {\cal O}_1 f \right)_{, \varphi} + \cos \theta \, {\cal O}_3 f &=& 0 \hspace{1cm} \mbox{(I2)} \, .
\label{id2}
\end{eqnarray}
These identities can also be directly checked with the definitions (\ref{opO}) of the operators ${\cal O}_i$.

The identities (I1)-(I2) can be differentiated to generate identities of higher order.
A first useful identity is obtained from $\sin\theta\mbox{(I1)}_{,\theta}-\mbox{(I2)}_{,\varphi}=0$:
\begin{equation}
\csc^2 \theta \left(
\left( \sin^2 \theta \left( {\cal O}_2 f \right)_{, \theta} \right)_{, \theta} + \left( {\cal O}_1 f \right)_{, \varphi , \varphi} - 2 \left( \sin \theta \, \left( {\cal O}_3 f \right)_{, \varphi} \right)_{, \theta}
\right)
- \cot \theta \, \left( {\cal O}_1 f \right)_{, \theta} + 2 \, {\cal O}_1 f = \Delta' f \, .
\label{id3} 
\end{equation}

A second useful identity is obtained from $\mbox{(I1)}_{, \varphi}+\sin\theta\mbox{(I2)}_{,\theta}=0$:
\begin{equation}
\csc^2 \theta \left(
\left( \sin^2 \theta \left( {\cal O}_3 f \right)_{, \theta} \right)_{, \theta} - \left( {\cal O}_3 f \right)_{, \varphi , \varphi} + \left( \sin \theta \left( \left( {\cal O}_2-{\cal O}_1 \right) f \right)_{, \varphi} \right)_{, \theta}
\right)
+ \cot \theta \left( {\cal O}_3 f \right)_{, \theta} - 2 \, {\cal O}_3 f = 0 \, .
\label{id4} 
\end{equation}

\subsection{No degree one in operators ${\cal A}$ and ${\cal B}$}
\label{Adegree1}

We want to prove that the operators ${\cal A}$ and ${\cal B}$ defined by equations (\ref{defA}) and (\ref{defB}) do not have any degree-one term in their spherical harmonic expansion:
\begin{eqnarray}
\int_{\cal S} d\omega \; {\cal A}(a \,; b) \, Y_{1p}^* &=& 0 \hspace{1cm} (p=-1,0,1) \, ,
\label{idA}
\\
\int_{\cal S} d\omega \; {\cal B}(a \,; b) \, Y_{1p}^* &=& 0 \hspace{1cm} (p=-1,0,1) \, ,
\label{idB}
\end{eqnarray}
where $(a,b)$ are arbitrary scalar functions on the sphere, $d\omega=\sin \theta \, d\theta \, d\varphi$ and the integral is taken over the whole spherical surface.

This property is not a straightforward consequence of constructing ${\cal A}$ and ${\cal B}$ with ${\cal D}_{ij}$ as a building block.
Although ${\cal A}$ and ${\cal B}$ can be factored into terms without degree one (such as ${\cal D}_{ij}a$ or $\Delta'a$), the product of the factors may contain degree-one terms in its spherical harmonic expansion.

Without loss of generality, we can prove the above identities with the arguments $(a,b)$ being spherical harmonics of given degree and order.
The general result is then obtained by superposition.
Let $a$ and $b$ be spherical harmonics of order $m$ and $n$: $a\sim e^{im\varphi}$ and $b\sim e^{in\varphi}$ (we will not use their harmonic degree in the proof).
All derivatives with respect to $\varphi$ in the operators ${\cal A}$ and ${\cal B}$ can then be replaced with the rules $a_{,\varphi}\rightarrow i m a$ and $b_{,\varphi}\rightarrow i n b$.
The integral over $\varphi$ in equations (\ref{idA})-(\ref{idB}) gives
\[
\int_0^{2\pi} d\varphi \; e^{i(m+n-p)\varphi} = 2 \pi \, \delta_{m+n-p,0} \, ,
\]
so that the integral is zero unless $n=p-m$.

First consider the case $p=0$ ($n=-m$), that is the projection on the zonal spherical harmonic of degree one.
We thus have to calculate
$\int_0^{\pi} d\theta \, A_0$ and $\int_0^{\pi} d\theta \, B_0$ with
\begin{eqnarray}
A_0 &\equiv& \sin \theta \, \cos \theta \, {\cal A}(a \,; b) \, ,
\nonumber \\
B_0 &\equiv& \sin \theta \, \cos \theta \, {\cal B}(a \,; b) \, .
\nonumber
\end{eqnarray}
The trick consists in rewriting the integrands as total derivatives:
\begin{eqnarray}
A_0 &=&
\Big(
\cos^2 \theta \, a_{, \theta} \, b_{, \theta} + \cot \theta \left( \sin^2 \theta - m^2 \right) (ab)_{, \theta} + \left( \sin^2 \theta + m^2 \, \csc^2 \theta \cos 2\theta \right) ab
\Big)_{,\theta} \, ,
\nonumber \\
B_0 &=&
-i \, m \Big(
\cos \theta \, a_{, \theta} \, b_{, \theta} + \sin \theta \, a \, b_{, \theta} - \csc \theta \cos^2 \theta \, a_{, \theta} \, b + \cos \theta \, a \, b_{, \theta , \theta}
\Big)_{,\theta} \, ,
\nonumber
\end{eqnarray}
The sought integrals are thus given by
\begin{eqnarray}
\int_0^{\pi} d\theta \, A_0 &=&
\Big[
a_{, \theta} \, b_{, \theta} - m^2 \, \cot \theta \,  (ab)_{, \theta} + m^2 \, \csc^2 \theta \, ab
\Big]_0^\pi \, ,
\nonumber \\
\int_0^{\pi} d\theta \, B_0 &=&
-i \, m \Big[
\cos \theta \, a_{, \theta} \, b_{, \theta} - \csc \theta \, a_{, \theta} \, b + \cos \theta \, a \, b_{, \theta , \theta}
\Big]_0^\pi \, ,
\nonumber
\end{eqnarray}
where we have dropped the terms containing at least one power of $\sin\theta$ which vanish at the limits;
we have also replaced $\cos^2\theta$ and $\cos2\theta$ by their value at the limits.
The remaining terms can be evaluated by recalling the dependence in $\sin\theta$ of the spherical harmonics:
$a = (\sin\theta)^{|m|} \, a_0$ and $b = (\sin\theta)^{|m|} \, b_0$, where $a_0$ and $b_0$ are polynomials in $\cos\theta$.
The only non-zero terms at the limits of the integrals are those for $|m|=1$, in which case we have at the limits:
$a_{, \theta}b_{, \theta}=a_0b_0$, $\cot\theta\,(ab)_{,\theta}=2a_0b_0$, $\csc^2\theta\,ab=a_0b_0$, $\csc\theta\,a_{,\theta}b=\cos\theta\,a_0b_0$, $ab_{,\theta,\theta}=0$.
However these terms cancel in the sums so that the integrals vanish for all $m$.
This completes the proof for the case $p=0$.

Now consider the case $p=\pm1$ ($n=-m\pm1$), that is the projections on the sectoral spherical harmonics of degree one.
We thus have to calculate
$\int_0^{\pi} d\theta \, A_{\pm1}$ and $\int_0^{\pi} d\theta \, B_{\pm1}$ with
\begin{eqnarray}
A_{\pm1} &\equiv& \sin^2 \theta \, {\cal A}(a \,; b) \, ,
\nonumber \\
B_{\pm1} &\equiv& \sin^2 \theta \, {\cal B}(a \,; b) \, .
\nonumber
\end{eqnarray}
We again write the integrands as total derivatives:
\begin{eqnarray}
A_{\pm1} &=&
\Big( \,
\sin \theta \cos \theta \, a_{, \theta} \, b_{, \theta} - \left( \sin\theta \cos\theta + m^2 \right) (ab)_{, \theta} - \left( \cos^2 \theta \mp 2m \right) a_{, \theta} \, b
\nonumber \\ &&
+ \, \sin^2 \theta \, a \, b_{, \theta} + 2m (m\mp1) \, \cot \theta \, ab
\, \Big)_{,\theta} \, ,
\nonumber \\
B_{\pm1} &=&
-i \, \Big( \,
(m\mp1) \sin \theta \, a_{, \theta} \, b_{, \theta} + m \sin \theta \, a \, b_{,\theta,\theta} - (m\mp1) \cos \theta \, a_{,\theta} \, b
\nonumber \\ &&
- \, m \cos  \theta \, a \, b_{, \theta} - m(1\mp m) \csc \theta \, ab
\, \Big)_{,\theta} \, .
\nonumber
\end{eqnarray}
The sought integrals are thus given by
\begin{eqnarray}
\int_0^{\pi} d\theta \, A_{\pm1} &=&
\Big[
- m^2 \, (ab)_{, \theta} - \left( 1 \mp 2m \right) a_{, \theta} \, b + 2m (m\mp1) \, \cot \theta \, ab
\, \Big]_0^\pi \, ,
\nonumber \\
\int_0^{\pi} d\theta \, B_{\pm1} &=&
i \Big[
(m\mp1) \cos \theta \, a_{,\theta} \, b + m \cos  \theta \, a \, b_{, \theta} + m(1\mp m) \csc \theta \, ab
\, \Big]_0^\pi \, ,
\nonumber
\end{eqnarray}
where we have dropped the terms containing at least one power of $\sin\theta$ and replaced $\cos^2\theta$ by its value at the limits.
The remaining terms can be evaluated as in the case $p=0$, but with $a = (\sin\theta)^{|m|} \, a_0$ and $b = (\sin\theta)^{|m\mp1|} \, b_0$.
All terms give zero at the limits of the integrals for all values of $m$.
This completes the proof for the case $p=\pm1$.
We have thus proven the identities (\ref{idA})-(\ref{idB}).

\small


\begin{thebibliography}{67}
\providecommand{\natexlab}[1]{#1}
\expandafter\ifx\csname urlstyle\endcsname\relax
  \providecommand{\doi}[1]{doi:\discretionary{}{}{}#1}\else
  \providecommand{\doi}{doi:\discretionary{}{}{}\begingroup
  \urlstyle{rm}\Url}\fi

\bibitem[{\textit{{Anderson} and {Smrekar}}(2006)}]{anderson06}
{Anderson}, F.~S., and S.~E. {Smrekar} (2006), {Global mapping of crustal and
  lithospheric thickness on Venus}, \textit{J.~Geophys.~Res.}, \textit{111},
  E08006, \doi{10.1029/2004JE002395}.
  
\bibitem[{\textit{{Arkani-Hamed}}(1998)}]{arkanihamed98}
{Arkani-Hamed}, J. (1998), {The lunar mascons revisited},
  \textit{J.~Geophys.~Res.}, \textit{103}, 3709--3739.

\bibitem[{\textit{{Arkani-Hamed}}(2000)}]{arkanihamed00}
{Arkani-Hamed}, J. (2000), {Strength of Martian lithosphere beneath large
  volcanoes}, \textit{J.~Geophys.~Res.}, \textit{105}, 26,713--26,732,
  \doi{10.1029/2000JE001267}.

\bibitem[{\textit{{Arkani-Hamed} and {Riendler}}(2002)}]{arkanihamed02}
{Arkani-Hamed}, J., and L.~{Riendler} (2002), {Stress differences in the
  Martian lithosphere: Constraints on the thermal state of Mars},
  \textit{J.~Geophys.~Res.}, \textit{107}, 2--1, \doi{10.1029/2002JE001851}.

\bibitem[{\textit{{Backus}}(1986)}]{backus86}
{Backus}, G. (1986), {Poloidal and toroidal fields in geomagnetic field
  modeling}, \textit{Rev.~Geophys.}, \textit{24}, 75--109.

\bibitem[{\textit{{Banerdt}}(1986)}]{banerdt86}
{Banerdt}, W.~B. (1986), {Support of long-wavelength loads on Venus and
  implications for internal structure}, \textit{J.~Geophys.~Res.}, \textit{91},
  403--419.

\bibitem[{\textit{{Banerdt} and {Golombek}}(2000)}]{banerdt00}
{Banerdt}, W.~B., and M.~P. {Golombek} (2000), {Tectonics of the Tharsis Region
  of Mars: Insights from MGS Topography and Gravity}, in \textit{Lunar and
  Planetary Institute Conference Abstracts}, p. 2038.

\bibitem[{\textit{{Banerdt} et~al.}(1982)\textit{{Banerdt}, {Saunders},
  {Phillips}, and {Sleep}}}]{banerdt82}
{Banerdt}, W.~B., R.~S. {Saunders}, R.~J. {Phillips}, and N.~H. {Sleep} (1982),
  {Thick shell tectonics on one-plate planets - Applications to Mars},
  \textit{J.~Geophys.~Res.}, \textit{87}, 9723--9733.

\bibitem[{\textit{{Banerdt} et~al.}(1992)\textit{{Banerdt}, {Golombek}, and
  {Tanaka}}}]{banerdt92}
{Banerdt}, W.~B., M.~P. {Golombek}, and K.~L. {Tanaka} (1992), {Stress and
  tectonics on Mars}, in \textit{Mars}, edited by {{Kieffer}, H.~H., {Jakosky},
  B.~M.,{Snyder}, C.~W. and {Matthews}, M.~S.}, pp. 249--297, University of
  Arizona Press, Tucson.

\bibitem[{\textit{{Belleguic} et~al.}(2005)\textit{{Belleguic}, {Lognonn{\'e}},
  and {Wieczorek}}}]{belleguic05}
{Belleguic}, V., P.~{Lognonn{\'e}}, and M.~{Wieczorek} (2005), {Constraints on
  the Martian lithosphere from gravity and topography data},
  \textit{J.~Geophys.~Res.}, \textit{110}, E11005, \doi{10.1029/2005JE002437}.

\bibitem[{\textit{{Blewitt}}(2003)}]{blewitt03}
{Blewitt}, G. (2003), {Self-consistency in reference frames, geocenter
  definition, and surface loading of the solid Earth},
  \textit{J.~Geophys.~Res.}, \textit{108}, 2103, \doi{10.1029/2002JB002082}.

\bibitem[{\textit{{Brotchie}}(1971)}]{brotchie71}
{Brotchie}, J.~F. (1971), {Flexure of a liquid-filled spherical shell in a
  radial gravity field}, \textit{Mod. Geol.}, \textit{3}, 15--23.

\bibitem[{\textit{{Brotchie} and {Silvester}}(1969)}]{brotchie69}
{Brotchie}, J.~F., and R.~{Silvester} (1969), {On crustal flexure},
  \textit{J.~Geophys.~Res.}, \textit{74}, 5249--5252.

\bibitem[{\textit{{Comer} et~al.}(1985)\textit{{Comer}, {Solomon}, and
  {Head}}}]{comer85}
{Comer}, R.~P., S.~C. {Solomon}, and J.~W. {Head} (1985), {Mars - Thickness of
  the lithosphere from the tectonic response to volcanic loads}, \textit{Rev.
  of Geophys.}, \textit{23}, 61--92.

\bibitem[{\textit{{Crosby} and {McKenzie}}(2005)}]{crosby05}
{Crosby}, A., and D.~{McKenzie} (2005), {Measurements of the elastic thickness
  under ancient lunar terrain}, \textit{Icarus}, \textit{173}, 100--107,
  \doi{10.1016/j.icarus.2004.07.017}.

\bibitem[{\textit{{Farrell}}(1972)}]{farrell72}
{Farrell}, W.~E. (1972), {Deformation of the Earth by Surface Loads},
  \textit{Rev. Geophys. Space Phys.}, \textit{10}, 761--797.

\bibitem[{\textit{{Forte} and {Peltier}}(1987)}]{forte87}
{Forte}, A.~M., and W.~R. {Peltier} (1987), {Plate tectonics and aspherical
  earth structure: The importance of poloidal-toroidal coupling},
  \textit{J.~Geophys.~Res.}, \textit{92}, 3645--3680.

\bibitem[{\textit{{Greff-Lefftz} and {Legros}}(1997)}]{greff97}
{Greff-Lefftz}, M., and L.~{Legros} (1997), {Some remarks about the degree-one
  deformation of the Earth}, \textit{Geophys.~J.~Int.}, \textit{131}, 699--723,
  \doi{10.1111/j.1365-246X.1997.tb06607.x}.

\bibitem[{\textit{{Hall} et~al.}(1986)\textit{{Hall}, {Solomon}, and
  {Head}}}]{hall86}
{Hall}, J.~L., S.~C. {Solomon}, and J.~W. {Head} (1986), {Elysium region, Mars
  - Tests of lithospheric loading models for the formation of tectonic
  features}, \textit{J.~Geophys.~Res.}, \textit{91}, 11,377--11,392.

\bibitem[{\textit{{Janes} and {Melosh}}(1990)}]{janes90}
{Janes}, D.~M., and H.~J. {Melosh} (1990), {Tectonics of planetary loading - A
  general model and results}, \textit{J.~Geophys.~Res.}, \textit{95},
  21,345--21,355.

\bibitem[{\textit{{Janle} and {Jannsen}}(1986)}]{janle86}
{Janle}, P., and D.~{Jannsen} (1986), {Isostatic gravity and elastic bending
  models of Olympus Mons, Mars}, \textit{Annales Geophysicae}, \textit{4},
  537--546.

\bibitem[{\textit{{Janle} and {Jannsen}}(1988)}]{janle88}
{Janle}, P., and D.~{Jannsen} (1988), {Tepev Mons on Venus: Morphology and
  Elastic Bending Models}, \textit{Earth Moon Planets}, \textit{41}, 127--139.

\bibitem[{\textit{{Johnson} and {Sandwell}}(1994)}]{johnson94}
{Johnson}, C.~L., and D.~T. {Sandwell} (1994), {Lithospheric flexure on Venus},
  \textit{Geophys.~J.~Int.}, \textit{119}, 627--647,
  \doi{10.1111/j.1365-246X.1994.tb00146.x}.

\bibitem[{\textit{{Johnson} et~al.}(2000)\textit{{Johnson}, {Solomon}, {Head},
  {Phillips}, {Smith}, and {Zuber}}}]{johnson00}
{Johnson}, C.~L., S.~C. {Solomon}, J.~W. {Head}, R.~J. {Phillips}, D.~E.
  {Smith}, and M.~T. {Zuber} (2000), {Lithospheric Loading by the Northern
  Polar Cap on Mars}, \textit{Icarus}, \textit{144}, 313--328,
  \doi{10.1006/icar.1999.6310}.

\bibitem[{\textit{{Kirby} and {Swain}}(2004)}]{kirby04}
{Kirby}, J.~F., and C.~J. {Swain} (2004), {Global and local isostatic coherence
  from the wavelet transform}, \textit{Geophys.~Res.~Lett.}, \textit{31},
  L24608, \doi{10.1029/2004GL021569}.

\bibitem[{\textit{{Kraus}}(1967)}]{kraus}
{Kraus}, H. (1967), \textit{{Thin Elastic Shells}}, John Wiley, New York.

\bibitem[{\textit{{Latychev} et~al.}(2005)\textit{{Latychev}, {Mitrovica},
  {Tamisiea}, {Tromp}, and {Moucha}}}]{latychev05}
{Latychev}, K., J.~X. {Mitrovica}, M.~E. {Tamisiea}, J.~{Tromp}, and
  R.~{Moucha} (2005), {Influence of lithospheric thickness variations on 3-D
  crustal velocities due to glacial isostatic adjustment},
  \textit{Geophys.~Res.~Lett.}, \textit{32}, L01304,
  \doi{10.1029/2004GL021454}.

\bibitem[{\textit{{Lawrence} and {Phillips}}(2003)}]{lawrence03}
{Lawrence}, K.~P., and R.~J. {Phillips} (2003), {Gravity/topography admittance
  inversion on Venus using niching genetic algorithms},
  \textit{Geophys.~Res.~Lett.}, \textit{30}, 1--1, \doi{10.1029/2003GL017515}.

\bibitem[{\textit{{Love}}(1944)}]{love}
{Love}, A.~E.~H. (1944), \textit{{A treatise on the mathematical theory of
  elasticity}}, 4th edition, Dover, New York.

\bibitem[{\textit{{Lowry} and {Zhong}}(2003)}]{lowry03}
{Lowry}, A.~R., and S.~{Zhong} (2003), {Surface versus internal loading of the
  Tharsis rise, Mars}, \textit{J.~Geophys.~Res.}, \textit{108}, 5099,
  \doi{10.1029/2003JE002111}.

\bibitem[{\textit{{McGovern} et~al.}(2002)\textit{{McGovern}, {Solomon},
  {Smith}, {Zuber}, {Simons}, {Wieczorek}, {Phillips}, {Neumann}, {Aharonson},
  and {Head}}}]{mcgovern02}
{McGovern}, P.~J., S.~C. {Solomon}, D.~E. {Smith}, M.~T. {Zuber}, M.~{Simons},
  M.~A. {Wieczorek}, R.~J. {Phillips}, G.~A. {Neumann}, O.~{Aharonson}, and
  J.~W. {Head} (2002), {Localized gravity/topography admittance and correlation
  spectra on Mars: Implications for regional and global evolution},
  \textit{J.~Geophys.~Res.}, \textit{107}, 5136, \doi{10.1029/2002JE001854}.

\bibitem[{\textit{{McGovern} et~al.}(2004)\textit{{McGovern}, {Solomon},
  {Smith}, {Zuber}, {Simons}, {Wieczorek}, {Phillips}, {Neumann}, {Aharonson},
  and {Head}}}]{mcgovern04}
{McGovern}, P.~J., S.~C. {Solomon}, D.~E. {Smith}, M.~T. {Zuber}, M.~{Simons},
  M.~A. {Wieczorek}, R.~J. {Phillips}, G.~A. {Neumann}, O.~{Aharonson}, and
  J.~W. {Head} (2004), {Correction to ``Localized gravity/topography admittance
  and correlation spectra on Mars: Implications for regional and global
  evolution''}, \textit{J.~Geophys.~Res.}, \textit{109}, E07007,
  \doi{10.1029/2004JE002286}.

\bibitem[{\textit{{McKenzie} et~al.}(2002)\textit{{McKenzie}, {Barnett}, and
  {Yuan}}}]{mckenzie02}
{McKenzie}, D., D.~N. {Barnett}, and D.-N. {Yuan} (2002), {The relationship
  between Martian gravity and topography}, \textit{Earth~Planet.~Sci.~Lett.},
  \textit{195}, 1--16.

\bibitem[{\textit{{M{\'e}tivier} et~al.}(2006)\textit{{M{\'e}tivier},
  {Greff-Lefftz}, and {Diament}}}]{metivier06}
{M{\'e}tivier}, L., M.~{Greff-Lefftz}, and M.~{Diament} (2006), {Mantle lateral
  variations and elastogravitational deformations - I. Numerical modelling},
  \textit{Geophys.~J.~Int.}, \textit{167}, 1060--1076,
  \doi{10.1111/j.1365-246X.2006.03159.x}.

\bibitem[{\textit{{Novozhilov}}(1964)}]{novozhilov}
{Novozhilov}, V.~V. (1964), \textit{{Thin shell theory}}, 2nd edition,
  Noordhoff, Groningen.

\bibitem[{\textit{{P{\'e}rez-Gussiny{\'e}}
  et~al.}(2004)\textit{{P{\'e}rez-Gussiny{\'e}}, {Lowry}, {Watts}, and
  {Velicogna}}}]{perezgussinye04}
{P{\'e}rez-Gussiny{\'e}}, M., A.~R. {Lowry}, A.~B. {Watts}, and I.~{Velicogna}
  (2004), {On the recovery of effective elastic thickness using spectral
  methods: Examples from synthetic data and from the Fennoscandian Shield},
  \textit{J.~Geophys.~Res.}, \textit{109}, B10409, \doi{10.1029/2003JB002788}.

\bibitem[{\textit{{Phillips} et~al.}(2001)\textit{{Phillips}, {Zuber},
  {Solomon}, {Golombek}, {Jakosky}, {Banerdt}, {Smith}, {Williams}, {Hynek},
  {Aharonson}, and {Hauck}}}]{phillips01}
{Phillips}, R.~J., M.~T. {Zuber}, S.~C. {Solomon}, M.~P. {Golombek}, B.~M.
  {Jakosky}, W.~B. {Banerdt}, D.~E. {Smith}, R.~M.~E. {Williams}, B.~M.
  {Hynek}, O.~{Aharonson}, and S.~A. {Hauck} (2001), {Ancient Geodynamics and
  Global-Scale Hydrology on Mars}, \textit{Science}, \textit{291}, 2587--2591,
  \doi{10.1126/science.1058701}.

\bibitem[{\textit{{Ranalli}}(1987)}]{ranalli}
{Ranalli}, G. (1987), \textit{{Rheology of the Earth}}, Allen and Unwin,
  Boston.

\bibitem[{\textit{{Ronchi} et~al.}(1996)\textit{{Ronchi}, {Iacono}, and
  Paolucci}}]{ronchi96}
{Ronchi}, C., R.~{Iacono}, and P.~S. Paolucci (1996), {The cubed sphere: a new
  method for the solution of partial differential equations in spherical
  geometry}, \textit{J. Comput. Phys.}, \textit{124}, 93--114.

\bibitem[{\textit{{Sandwell}}(1984)}]{sandwell84}
{Sandwell}, D.~T. (1984), {Thermomechanical evolution of oceanic fracture
  zones}, \textit{J.~Geophys.~Res.}, \textit{89}, 11,401--11,413.

\bibitem[{\textit{{Sandwell} et~al.}(1997)\textit{{Sandwell}, {Johnson},
  {Bilotti}, and {Suppe}}}]{sandwell97}
{Sandwell}, D.~T., C.~L. {Johnson}, F.~{Bilotti}, and J.~{Suppe} (1997),
  {Driving Forces for Limited Tectonics on Venus}, \textit{Icarus},
  \textit{129}, 232--244, \doi{10.1006/icar.1997.5721}.

\bibitem[{\textit{{Searls} et~al.}(2006)\textit{{Searls}, {Banerdt}, and
  {Phillips}}}]{searls06}
{Searls}, M.~L., W.~B. {Banerdt}, and R.~J. {Phillips} (2006), {Utopia and
  Hellas basins, Mars: Twins separated at birth}, \textit{J.~Geophys.~Res.},
  \textit{111}, E08005, \doi{10.1029/2005JE002666}.

\bibitem[{\textit{{Simons} et~al.}(1997)\textit{{Simons}, {Solomon}, and
  {Hager}}}]{simons97}
{Simons}, M., S.~C. {Solomon}, and B.~H. {Hager} (1997), {Localization of
  gravity and topography: Constraints on the tectonics and mantle dynamics of
  Venus}, \textit{Geophys.~J.~Int.}, \textit{131}, 24--44,
  \doi{10.1111/j.1365-246X.1997.tb00593.x}.

\bibitem[{\textit{{Sleep} and {Phillips}}(1985)}]{sleep85}
{Sleep}, N.~H., and R.~J. {Phillips} (1985), {Gravity and lithospheric stress
  on the terrestrial planets with reference to the Tharsis region of Mars},
  \textit{J.~Geophys.~Res.}, \textit{90}, 4469--4489.

\bibitem[{\textit{{Sokolnikoff}}(1956)}]{sokolnikoff}
{Sokolnikoff}, I.~S. (1956), \textit{{Mathematical Theory of Elasticity}},
  McGraw-Hill, New York.

\bibitem[{\textit{{Solomon}}(1978)}]{solomon78}
{Solomon}, S.~C. (1978), {On volcanism and thermal tectonics on one-plate
  planets}, \textit{Geophys.~Res.~Lett.}, \textit{5}, 461--464.

\bibitem[{\textit{{Solomon} and {Head}}(1979)}]{solomon79}
{Solomon}, S.~C., and J.~W. {Head} (1979), {Vertical movement in mare basins -
  Relation to mare emplacement, basin tectonics, and lunar thermal history},
  \textit{J.~Geophys.~Res.}, \textit{84}, 1667--1682.

\bibitem[{\textit{{Solomon} and {Head}}(1982)}]{solomon82}
{Solomon}, S.~C., and J.~W. {Head} (1982), {Evolution of the Tharsis province
  of Mars: The importance of heterogeneous lithospheric thickness and volcanic
  construction}, \textit{J.~Geophys.~Res.}, \textit{87}, 9755--9774.

\bibitem[{\textit{{Solomon} and {Head}}(1990)}]{solomon90}
{Solomon}, S.~C., and J.~W. {Head} (1990), {Heterogeneities in the thickness of
  the elastic lithosphere of Mars: Constraints on heat flow and internal
  dynamics}, \textit{J.~Geophys.~Res.}, \textit{95}, 11,073--11,083.

\bibitem[{\textit{{Stark} et~al.}(2003)\textit{{Stark}, {Stewart}, and
  {Ebinger}}}]{stark03}
{Stark}, C.~P., J.~{Stewart}, and C.~J. {Ebinger} (2003), {Wavelet transform
  mapping of effective elastic thickness and plate loading: Validation using
  synthetic data and application to the study of southern African tectonics},
  \textit{J.~Geophys.~Res.}, \textit{108}, 2558, \doi{10.1029/2001JB000609}.

\bibitem[{\textit{{Stewart} and {Watts}}(1997)}]{stewart97}
{Stewart}, J., and A.~B. {Watts} (1997), {Gravity anomalies and spatial
  variations of flexural rigidity at mountain ranges},
  \textit{J.~Geophys.~Res.}, \textit{102}, 5327--5352, \doi{10.1029/96JB03664}.

\bibitem[{\textit{{Sugano} and {Heki}}(2004)}]{sugano04}
{Sugano}, T., and K.~{Heki} (2004), {Isostasy of the Moon from high-resolution
  gravity and topography data: Implication for its thermal history},
  \textit{Geophys.~Res.~Lett.}, \textit{31}, L24703,
  \doi{10.1029/2004GL022059}.

\bibitem[{\textit{{Synge} and {Schild}}(1978)}]{synge}
{Synge}, J.~L., and J.~N. {Schild} (1978), \textit{{Tensor calculus}}, Dover,
  New York.

\bibitem[{\textit{{Tanimoto}}(1998)}]{tanimoto98}
{Tanimoto}, T. (1998), {State of stress within a bending spherical shell and
  its implications for subducting lithosphere}, \textit{Geophys. J. Int.},
  \textit{134}, 199--206, \doi{10.1046/j.1365-246x.1998.00554.x}.

\bibitem[{\textit{{Thurber} and {Toks{\"o}z}}(1978)}]{thurber78}
{Thurber}, C.~H., and M.~N. {Toks{\"o}z} (1978), {Martian lithospheric
  thickness from elastic flexure theory}, \textit{Geophys.~Res.~Lett.},
  \textit{5}, 977--980.

\bibitem[{\textit{{Timoshenko} and {Woinowsky-Krieger}}(1964)}]{timoshenkoW}
{Timoshenko}, S., and S.~{Woinowsky-Krieger} (1964), \textit{{Theory of Plates
  and Shells (2nd edition)}}, McGraw-Hill, New York.

\bibitem[{\textit{{Turcotte} et~al.}(1981)\textit{{Turcotte}, {Willemann},
  {Haxby}, and {Norberry}}}]{turcotte81}
{Turcotte}, D.~L., R.~J. {Willemann}, W.~F. {Haxby}, and J.~{Norberry} (1981),
  {Role of membrane stresses in the support of planetary topography},
  \textit{J.~Geophys.~Res.}, \textit{86}, 3951--3959.

\bibitem[{\textit{{Turcotte} et~al.}(2002)\textit{{Turcotte}, {Shcherbakov},
  {Malamud}, and {Kucinskas}}}]{turcotte02}
{Turcotte}, D.~L., R.~{Shcherbakov}, B.~D. {Malamud}, and A.~B. {Kucinskas}
  (2002), {Is the Martian crust also the Martian elastic lithosphere?},
  \textit{J.~Geophys.~Res.}, \textit{107}, 5091, \doi{10.1029/2001JE001594}.

\bibitem[{\textit{{Vlasov}}(1964)}]{vlasov}
{Vlasov}, V.~Z. (1964), \textit{{General theory of shells and its applications
  in engineering}}, NASA Tech. Trans. TT F-99, 1-886.

\bibitem[{\textit{{Wang} and {Wu}}(2006)}]{wang06}
{Wang}, H., and P.~{Wu} (2006), {Effects of lateral variations in lithospheric
  thickness and mantle viscosity on glacially induced surface motion on a
  spherical, self-gravitating Maxwell Earth},
  \textit{Earth~Planet.~Sci.~Lett.}, \textit{244}, 576--589,
  \doi{10.1016/j.epsl.2006.02.026}.

\bibitem[{\textit{{Wieczorek}}(2007)}]{wieczorek06}
{Wieczorek}, M.~A. (2007), {The gravity and topography of the terrestrial
  planets}, in \textit{Treatise on Geophysics (in press)}, edited by
  {{Schubert}, G.}, Elsevier, Amterdam.

\bibitem[{\textit{{Wieczorek} and {Simons}}(2005)}]{wieczorek05}
{Wieczorek}, M.~A., and F.~J. {Simons} (2005), {Localized spectral analysis on
  the sphere}, \textit{Geophys.~J.~Int.}, \textit{162}, 655--675,
  \doi{10.1111/j.1365-246X.2005.02687.x}.

\bibitem[{\textit{{Willemann} and {Turcotte}}(1982)}]{willemann82}
{Willemann}, R.~J., and D.~L. {Turcotte} (1982), {The role of lithospheric
  stress in the support of the Tharsis rise}, \textit{J.~Geophys.~Res.},
  \textit{87}, 9793--9801.

\bibitem[{\textit{{Zhong}}(2002)}]{zhong02}
{Zhong}, S. (2002), {Effects of lithosphere on the long-wavelength gravity
  anomalies and their implications for the formation of the Tharsis rise on
  Mars}, \textit{J.~Geophys.~Res.}, \textit{107}, 5054--+,
  \doi{10.1029/2001JE001589}.

\bibitem[{\textit{{Zhong} and {Roberts}}(2003)}]{zhong03}
{Zhong}, S., and J.~H. {Roberts} (2003), {On the support of the Tharsis Rise on
  Mars}, \textit{Earth~Planet.~Sci.~Lett.}, \textit{214}, 1--9,
  \doi{10.1016/S0012-821X(03)00384-4}.

\bibitem[{\textit{{Zhong} and {Zuber}}(2000)}]{zhong00}
{Zhong}, S., and M.~T. {Zuber} (2000), {Long-wavelength topographic relaxation
  for self-gravitating planets and implications for the time-dependent
  compensation of surface topography}, \textit{J.~Geophys.~Res.}, \textit{105},
  4153--4164, \doi{10.1029/1999JE001075}.

\bibitem[{\textit{{Zhong} et~al.}(2003)\textit{{Zhong}, {Paulson}, and
  {Wahr}}}]{zhongpau03}
{Zhong}, S., A.~{Paulson}, and J.~{Wahr} (2003), {Three-dimensional
  finite-element modelling of Earth's viscoelastic deformation: effects of
  lateral variations in lithospheric thickness}, \textit{Geophys.~J.~Int.},
  \textit{155}, 679--695, \doi{10.1046/j.1365-246X.2003.02084.x}.

\end{thebibliography}
\end{document}